\documentclass[aps,pre,reprint,longbibliography,amsmath,amssymb,groupedaddress]{revtex4-1}

\pdfoutput=1

\newcommand{\vect}[1]{\boldsymbol{#1}}

\newcommand{\dtheta}{\partial_{\theta}}
\newcommand{\dphi}{\partial_{\phi}}

\newcommand{\ephi}{\hat{\vect{e}}_{\phi}}
\newcommand{\etheta}{\hat{\vect{e}}_{\theta}}

\usepackage{hyperref}
\usepackage{graphicx}
\usepackage{physics}
\usepackage{bm}
\begin{document}

\title{Vortex formation and dynamics of defects in shells of active nematics}

\author{Diana Khoromskaia}
\author{Gareth P. Alexander}
\affiliation{Department of Physics and Centre for Complexity Science, University of Warwick, Coventry CV4 7AL, United Kingdom}

\date{\today}

\begin{abstract}
We present a hydrodynamic model for a thin spherical shell of active nematic liquid crystal with an arbitrary configuration of defects. The active flows generated by defects in the director lead to the formation of stable vortices, analogous to those seen in confined systems in flat geometries, which generate an effective dynamics for four $+1/2$ defects that reproduces the tetrahedral to planar oscillations observed in experiments. As the activity is increased and the vortices become stronger, the defects are drawn more tightly into pairs, rotating about antipodal points. We extend this situation to also describe the dynamics of other configurations of defects. For example, two $+1$ defects are found to attract or repel according to the local geometric character of the director field around them and the extensile or contractile nature of the material, while additional pairs of opposite charge defects can give rise to flow states containing more than two vortices. Finally, we describe the generic relationship between defects in the orientation and singular points of the flow, and suggest implications for the three-dimensional nature of the flow and deformation in the shape of the shell.
\end{abstract}

\maketitle

\section{Introduction}
Active liquid crystals \cite{2015NatPh..11..111P,Marchetti:2013bp,Ramaswamy:2010bf} (ALCs) have proved successful as a paradigm for living systems on the microscale, providing insight into processes like cell motility\cite{Kruse:2006gq,2015NatCo...6E5420T,2012PNAS..10912381T} and division\cite{Salbreux:2009fp,Turlier:2014hq,Brugues:2014bp}, development of cell shapes  \cite{Salbreux:2007fc,CallanJones:2016fe}, and growth of cell colonies \cite{Doostmohammadi:2016iq}. Certain fundamental motifs have been developed such as the instability of uniformly aligned states, the emergence of spontaneous flows, the creation and self-propulsion of topological defects and the shear-thinning character of extensile gels. When ALCs are confined to a circular geometry a prominent feature is the emergence of stable flow vortices. Confinement gives rise to a single vortex state in dense bacterial suspensions \cite{Wioland:2013jm,Wioland:2016kv}, active nematic suspensions \cite{2015arXiv151103880G,Woodhouse:2012cl} and monolayers of migrating cells \cite{Segerer:2015js,MohammadNejad:2014he}. Circulatory flows are also characteristic of cytoplasmic streaming \cite{2012PNAS..10912381T,AWhitfield:2014in,Kumar:2014kv,Woodhouse:2012cl}. When the system size is increased such vortices become unstable \cite{AditiSimha:2002eg} and turbulent flows develop, a prevalent feature in bulk active fluids \cite{Sanchez:2013gt,Dunkel:2013bm,Giomi:2015fu}. In active systems with high frictional dissipation stable vortices can also arise in the absence of spatial confinement \cite{Adamer:2016bd,Schaller:2010cq,Sumino:2012dw}. Recent experiments by Keber {\it et al.} \cite{Keber:2014fh} are realisations of a different type of confined geometry, in which the ALC adheres to the surface of a vesicle. Four half-integer defects form in the orientation of the microtubule-based extensile active nematic and are found to be in steady motion, oscillating between tetrahedral and planar configurations. A variety of other states is observed, like defect-associated membrane protrusions, two vortex defects in smaller spherical vesicles and two aster defects in spindle-like vesicles with stiffer microtubules. Here, we develop an active hydrodynamic model for an ALC confined to a spherical shell and show that the dynamics of this system is also characterised by the formation of vortices, which reproduces the defect motion from experiments.

Defects in the director are unavoidable on the sphere~\cite{Hopf}. In a typical situation there are four, all of strength $+1/2$, which are known to self-propel in active liquid crystals \cite{Giomi:2014ha}. This motivates a minimal description of their motion as a point particle dynamics, and such a model was shown to reproduce the main experimental observations~\cite{Keber:2014fh}. We extend this to a hydrodynamic model, in the confined geometry of a spherical shell, and show that the dynamics is characterised by the formation of two stable counterrotating vortices, one in each hemisphere, paralleling the vortex formation seen in other types of confinement~\cite{Woodhouse:2012cl,Wioland:2013jm,Sknepnek:2015gm}. A minimal hydrodynamic model takes the positions of defects to construct a profile for the director over the entire sphere, whose associated active flows advect the defects to yield a self-consistent dynamics. Tetrahedral to planar oscillations of four +1/2 defects are also obtained with this model. As the activity is increased the two vortices become more pronounced and the pair of $+1/2$ defects within each are pulled closer together in an effective attraction of like-charge defects. These oscillations appear at a finite threshold of the activity, below which the defects form static configurations of distorted tetrahedra. Linear stability analysis captures the mode of deformation and the threshold for defect motion. 

Just as there are defects in the director field there are also vortices and stagnation points in the flow field ~\cite{Hopf}. There is a one-way relationship that assigns to a defect in the orientation a flow singularity whose winding number depends only on the defect's topological strength. The oscillations of four half-defects are found to be stable against additional half-defect pairs created randomly in larger shells. If the defects are instead induced at specific positions, it is possible to generate more complex, metastable flow vortex configurations. The dynamics of polar configurations with only integer strength defects is similar and we find  attraction of pairs of aster-like +1 defects in extensile active nematic shells, but repulsion for vortex-like defects. The speed of defects in the polar case is shown to have different scaling than for nematic shells, in particular the type of motion does not depend on the radius in the former case whereas it does in the latter.

\section{Model}
We consider an active nematic in a thin spherical shell of thickness $h_0$ and inner radius $R$, with $h_0/R \ll 1$.  The three-dimensional flow $\vect{u}=(u_r, \vect{u}_{\perp})$ in the shell is driven by gradients in the active stresses and can be found as the solution of the generalised Stokes and continuity equations, $- \nabla p + \mu \Delta \vect{u} + \nabla \cdot \vect{\sigma} =  \vect{0} $ and $\nabla \cdot \vect{u}= 0$, where $p$ is the pressure and $\mu$ the viscosity.  The active stress $\vect{\sigma}^a = -\sigma_0 \left( \vect{P}\vect{P} - \frac{1}{3} \mathbb{I}\right)$ is extensile throughout this paper, $\sigma_0 > 0$, in order to relate with microtubule-based active nematics \cite{Sanchez:2013gt,Keber:2014fh}, although we comment on the contractile case at the end. If the polarisation $\vect{P}$ is specified one can solve for the active flow generated by it in a thin film approach \cite{1997RvMP...69..931O,TAKAGI:2010fi,Sankararaman:2009bx,Joanny:2012dx,Khoromskaia:2015ec},  decribed in Appendix \ref{app:flow}. We take the polarisation to be tangential throughout the shell thickness, $\vect{P} = \cos(\psi) \hat{\vect{e}}_{\theta}  + \sin(\psi)\hat{\vect{e}}_{\phi}$, and construct an explicit form from the positions of the defects. This can be done using stereographic projection from the complex plane,   $z(\theta,\phi) = R \cot(\theta/2)e^{i\phi}$. In the plane a nematic director $\vect{n}=\left(\cos \alpha , \sin \alpha \right)$ with $n_{\mathrm{def}}$ defects with topological strengths $m_j$ and positions $z_j=x_j+iy_j$ is given by $\alpha = \alpha_0 + \sum_j \mathrm{Im}\left( \ln (z-z_j)^{m_j} \right)$~\cite{Chaikin}, where the phase $\alpha_0 \in [0, \pi)$ parameterises whether the local geometry of the director around a defect is more splay-like or more bend-like. Finally, stereographic projection of $\vect{n}$ onto the sphere yields a polarisation field in the spherical shell via
\begin{equation}\label{eq:ch4:parametrisation}
\psi(\theta,\phi) = \phi - \alpha(\theta,\phi).
\end{equation}
Parametrised in this way, $\vect{P}$ is an exact minimiser of the elastic energy of a nematic on a sphere in the one-elastic-constant approximation \cite{Lubensky:1992bn}. Moreover, it consists only of those defects from which $\alpha$ is constructed explicitly, provided $\sum_j m_j =2$. 

\begin{figure}[t]
	\includegraphics[width=0.49\textwidth]{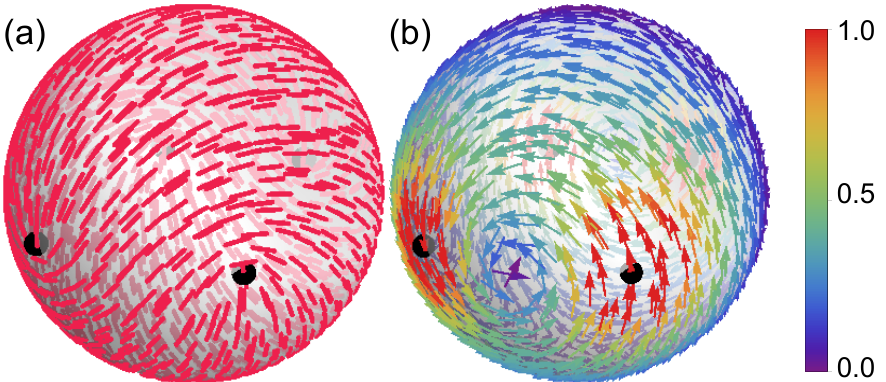}
	\caption{(a) Orientation field $\vect{P}$ with four +1/2 defects in a paired planar configuration. (b) Resulting tangential flow $\vect{u}_{\perp}$ from Eq.\ \eqref{eq:flow}, showing the typical two-vortex structure.  The flow magnitude is colour-coded and cut off in the vicinity of the defects. Displayed is the outer surface of the shell, $\tilde{r}=1$. \label{fig:fourdefects_example}}
\end{figure}

In dimensionless variables the tangential component of the flow then has the form
\begin{equation}\label{eq:flow}
\tilde{\vect{u}}_{\perp} = \tilde{\sigma}_0 f(\tilde{r}) \left(
\begin{matrix}
-\sin (2\psi) \dtheta \psi + \frac{\cos (2\psi)}{\sin \theta}\left(\cos \theta  + \dphi \psi  \right) \\
\cos (2\psi) \dtheta \psi + \frac{\sin (2 \psi)}{\sin\theta}\left(\cos \theta +\dphi \psi  \right) 
\end{matrix}\right),
\end{equation}
with the radial profile $f(\tilde{r})= \frac{\tilde{r}^2}{2}-\tilde{r}$, where $\tilde{r}\in [0,1]$ is the radial position within the shell in units of $h_0$. This solution corresponds to a no-slip inner surface and a vanishing tangential stress on the outer surface. Figure \ref{fig:fourdefects_example} gives an example of the director for four +1/2 defects in a planar configuration and the corresponding active flow given by Eq.\ \eqref{eq:flow}, which is seen to consist of two counterrotating vortices. This emergence of stable vortices is the germane feature of the active flows on spherical shells. 

\begin{figure*}[t]
	\centering
	\includegraphics[width=0.95\textwidth]{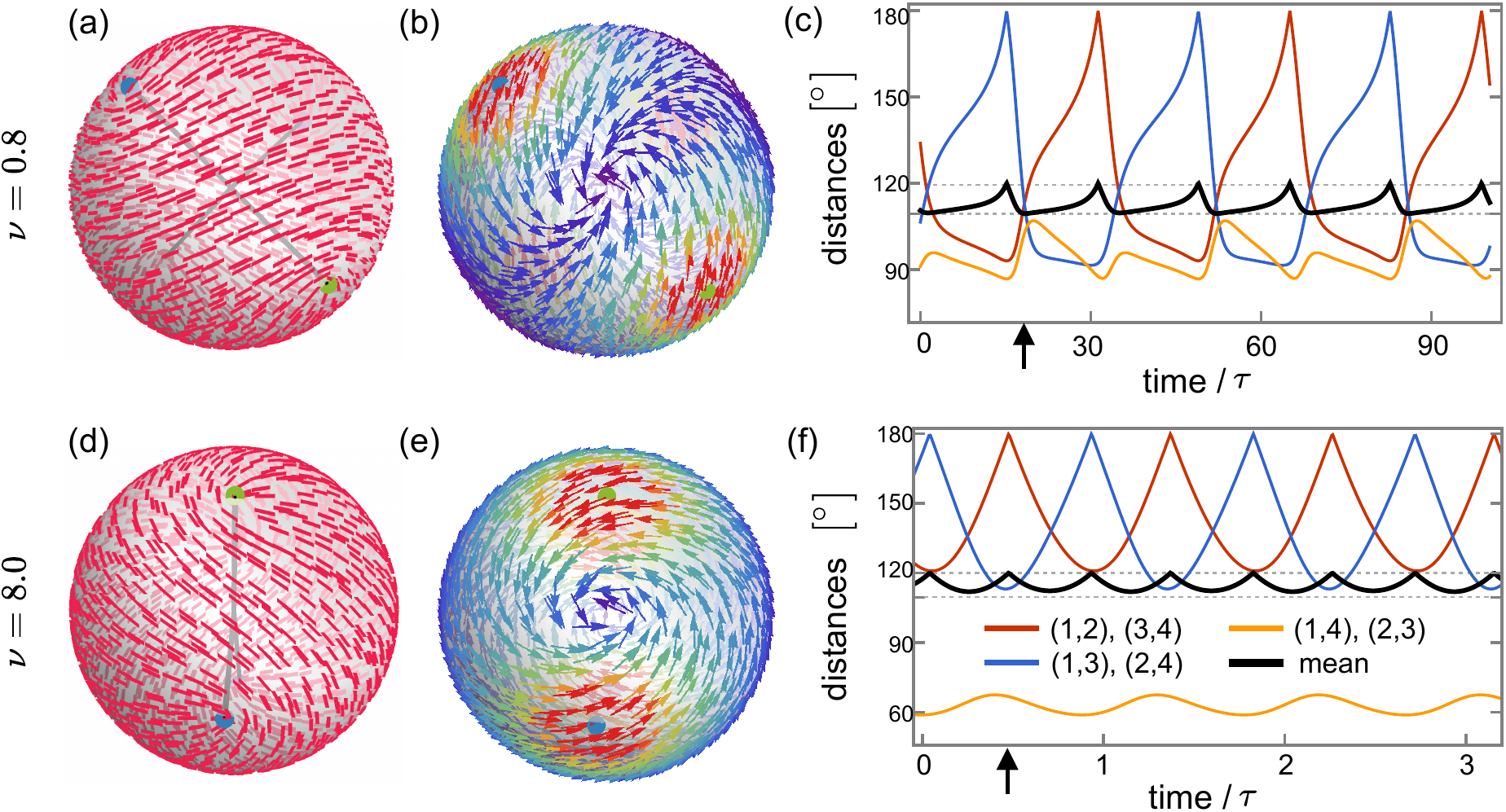}
	\caption{The dynamics of four +1/2 defects is chacterised by the formation of counterrotating flow vortices and defect pairs. (a)-(c): Tetrahedral-planar oscillations in intermediate activity regime ($\nu = 0.8$); Director (a) and flow (b) in tetrahedral configuration at time marked with arrow in (c). The flow vortex inbetween the paired defects in (b) has a sink-like component, which provides an effective attraction of defects and keeps them in pairs. (c) Pairwise and mean angular distances between defects; tetrahedral and planar configurations correspond to $109.5^\circ$ and $120^\circ$, respectively. (d)-(f): At higher activity ($\nu=8.0$) the two vortices become more pronounced and the defect pairs are tighter (see yellow line); Director (d) and flow (e) in planar configuration at time marked with arrow in (f); The tetrahedron is no longer approached as seen from pairwise distances in (f). In (c) and (f) $(i,j)$ denotes the distance between defects marked as $(i)$ and $(j)$ in Figure \ref{fig:parameterscan}.	\label{fig:fourdefects}}
\end{figure*}

The director dynamics is dominated by the motion of defects, when the orientational dynamics is rapid. In our approach the director is instantaneously given by the parameterisation above as the defect positions change. The defects are advected by the flow they create and we describe their motion in a point-particle description \cite{Giomi:2014ha,Pismen:2013ie}. Each defect moves due to the tangential component of the active flows, given by Eq.\ \eqref{eq:flow}, and due to standard nematic elasticity. The overdamped dynamical system for the defect positions $\vect{r}_k(t)$ is 
\begin{equation}\label{eq:dynamics}
\frac{\mathrm{d} \vect{r}_k(t)}{\mathrm{d} t} = \vect{u}_{k}^{\mathrm{def}}(t) + \frac{1}{\xi}\vect{F}_k (t) \, ,\quad k=1,.., n_{\mathrm{def}}.
\end{equation}
The resultant dynamics is similar to \cite{Keber:2014fh} except that here we obtain the advective flow $\vect{u}_{k}^{\mathrm{def}}$ from a self-consistent hydrodynamics in the spherical shell and generalise to an arbitrary collection of defects. The flow is divergent at the defect locations, therefore we introduce a cut-off and obtain the defect velocity as an average of the flow over a small circle $\gamma_k(s)$ centered at the defect 
\begin{equation}\label{eq:integral}
\vect{u}_{k}^{\mathrm{def}}(t) = \frac{1}{2\pi}\oint_{\gamma_k}\! \vect{u}_{\perp}(t) \mathrm{d}s.
\end{equation}
For the defect motion the flow is evaluated at the outer surface, where $f(\tilde{r}=1)=-1/2$. 
The circle $\gamma_k(s)= (\theta_k + \rho \cos(s), \phi_k +\rho\sin(s)/\sin(\theta_k))$, $s\in [0,2\pi]$, has the opening angle $\rho$, which can be associated with the core size $r_c$ of the defect through the relation
\begin{equation}
r_c = \rho R \,.
\end{equation}
The core size could be measured for a particular experimental system, for instance as the size of the region around a defect which is devoid of active nematogens. 

The elastic force $\vect{F}_k(t)$ provides attraction or repulsion of defects depending on their  topological strength, with an effective friction coefficient $\xi$ and elastic constant $K$ \cite{Vitelli:2006ba,Keber:2014fh} (see Appendix \ref{app:dynamics}). With the time scale of elastic relaxation $\tau=\xi R^2/K$ we define $\tilde{t}=t/\tau$ and equation \eqref{eq:dynamics} takes the form
\begin{eqnarray}\label{dynamicsnondim1}
\frac{\partial{\theta_k}}{\partial{\tilde{t}}} = & \hspace{-12mm} \frac{\tau}{R} u^{\mathrm{def}}_{k,\theta} +\frac{\tau}{\xi R} F_{k,\theta}, \\ \label{dynamicsnondim2}
\frac{\partial {\phi_k}}{\partial{\tilde{t}}}   = &\frac{1}{\sin\theta_k} \bigg(\frac{\tau}{R} u^{\mathrm{def}}_{k,\phi} + \frac{\tau}{\xi R} F_{k,\phi} \bigg).
\end{eqnarray}
This choice of time scale sets the scale of the elastic terms to $\tilde{K}=\tau K/\xi R^2=1$. This identifies the scaling of $\tau \abs{\vect{u}^{\mathrm{def}}}/R$ as the defining parameter for the defect dynamics, which represents the ratio of active to elastic effects and differs depending on the topological strength of the defect. Equations \eqref{dynamicsnondim1} and \eqref{dynamicsnondim2} are integrated numerically for different defect configurations using a standard Runge-Kutta method. 

\section{Results}
\subsection{Active flow at the defects}

In addition to the singularities in the director, the vortices in Fig. \ref{fig:fourdefects_example} (b) contain singularities in the flow field, about which the flow circulates. Such flow singularities are topologically required~\cite{Hopf} and can be generated at the locations of defects in the director. A general relationship between defects and flow singularities follows from evaluating \eqref{eq:flow} on the small circle $\gamma_k(s)$ and expanding in powers of $\rho$, the angular distance to the $k$-th defect (see Appendix \ref{app:flow}). We make use of the stereographic projection to write 
\begin{align}\nonumber
\tilde{u}(\rho)&=\tilde{u}_{\theta}+i\tilde{u}_{\phi} \\ \label{eq:expansion}
&= \frac{m_k}{\rho} e^{i(2m_k -1)s} e^{i2(1-m_k)\phi_k} e^{-i2w(z_k)} +\mathcal{O}(1)\,,
\end{align}
where $w(z_k) = \alpha_0 +m_k \pi+ \sum_{j\neq k} m_j \mathrm{Im}\left( \ln (z_k-z_j) \right)$. The dominant contribution to the flow $\tilde{\vect{u}}_{\perp}$ near the $k$-th defect diverges as $\sim 1/\rho$ and has the winding number 
\begin{equation}\label{eq:winding}
\mathcal{I}= 2m_k -1\,.
\end{equation}
Unit strength defects produce a  vortex-like ($\mathcal{I}= 1$) singularity in the flow, whose character is sink- or source-like according to whether the defect resembles an aster or a vortex, respectively. When there are two such defects, at antipodal positions, they generate two counterrotating vortices with no other flow singularities. On the other hand, simple stagnation points ($\mathcal{I}=-1$) cannot be created at defect locations. For half-integer defects relation \eqref{eq:winding} was shown in \cite{Giomi:2014ha,Pismen:2013ie} and the flow around a single spiral defect in active polar gels was studied in \cite{Kruse:2004il}.

In a typical situation the flow singularities at defects are not sufficient to generate a total winding of $+2$. This is most evident for four half-defects, as seen in Fig.\ \ref{fig:fourdefects_example} (b) where flow vortices form inbetween the defects, because for $m_k=1/2$ the flow is non-winding ($\mathcal{I}=0$). Instead, it is directed along the defect's symmetry axis
\begin{equation}
\tilde{u}^{+1/2} = \frac{1}{2\rho} e^{i(\phi_k-2w(z_k))}\,.
\end{equation}
For these defects, we approximate the advective flow $\vect{u}^{\mathrm{def}}$ in \eqref{eq:integral} by this well-defined flow direction and the magnitude
\begin{equation}\label{eq:activespeed}
|{\vect{u}^{\mathrm{def}}}| \sim \frac{U_0}{\rho} = \frac{h_0^2 \sigma_0}{r_c \mu}=:v_0\,,	
\end{equation}
where $U_0 = h_0^2 \sigma_0 /R \mu$ is the typical active flow magnitude in the thin film approach (see equation \eqref{eq:activescaling} in Appendix \ref{app:flow}) and we replaced $\rho=r_c/R$. The speed of $+1/2$ defects does not depend on the shell radius $R$, because they generate their own advection locally, where the defining length scales are  the core size $r_c$ and the shell thickness $h_0$. In equations \eqref{dynamicsnondim1} and \eqref{dynamicsnondim2} the scaling of the dimensionless advective term for a $+1/2$ defect is 
\begin{equation}\label{eq:ratio}
\frac{\tau}{R} |{\vect{u}^{\mathrm{def}}}| \sim \frac{\xi h_0^2 R \sigma_0}{K\mu r_c}=: \nu.
\end{equation}

The next term in the expansion \eqref{eq:expansion}, which is $\mathcal{O}(1)$, is non-winding only for $m_k=1$ (see equation \eqref{app:expansion} in Appendix \ref{app:flow}). Therefore, unit strength defects are advected with a flow $\sim U_0$, and the relevant parameter becomes
\begin{equation}\label{eq:ratio2}
\frac{\tau}{R} |\vect{u}^{\mathrm{def}}| \sim \frac{\xi h_0^2 \sigma_0}{K\mu}=: \nu^{(1)}\,.
\end{equation}
This predicts a different scaling of the defect dynamics in thin polar shells compared to nematic shells. In the former only integer strength defects are present and, notably, the type of motion does not depend on the radius. 

For all other defect types active advection scales at most as $\sim U_0 r_c/R$, which makes it negligible compared to the active motion of $+1/2$ defects. In particular, $-1/2$ defects can be approximated with $\vect{u}^{\mathrm{def}}=0$ in a collection of $\pm 1/2$ defects. 

\begin{figure}[t]
	\includegraphics[width=0.9\linewidth]{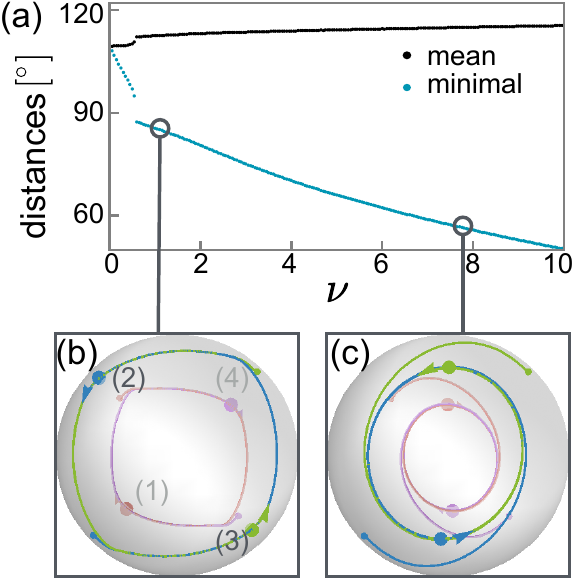}
	\caption{Mean and minimal angular distances between four +1/2 defects plotted against $\nu$, the ratio of active to elastic effects. The rapid change in both at $\nu^{*}\approx 0.7$ marks the transition to the dynamical regime, in which defects move on periodic orbits. The shape of these orbits changes smoothly with $\nu$: (b) square-like trajectories for intermediate activity, corresponding to the tetrahedral-planar oscillations, (c) ellipsoid orbits at higher activity, where defects in each pair have moved closer. In (b) and (c), small dots represent the initial tetrahedral configuration, big dots represent the defect positions at a time corresponding to plots in Figure \ref{fig:fourdefects} and arrows indicate the direction in which defects traverse the orbits.}
	\label{fig:parameterscan}
\end{figure}

\subsection{Four +1/2 defects}

In the minimal case of four $+1/2$ defects, the dynamics is determined by the parameter $\nu$, defined in \eqref{eq:ratio}. We increase $\nu$ through the activity $\sigma_0$, keeping all other parameters constant, in particular the radius, in order to fix the time scale $\tau$. The phase $\alpha_0$ also affects how the defects move. In the ranges $(0,\pi/4)$ and $(\pi/4,\pi/2)$ the dynamics is similar and we choose $\alpha_0=\pi/2-0.2$ for the examples in the plots. The marginal cases are discussed at the end of this Section. 

\begin{figure*}[th]
	\centering
	\includegraphics[width=0.99\linewidth]{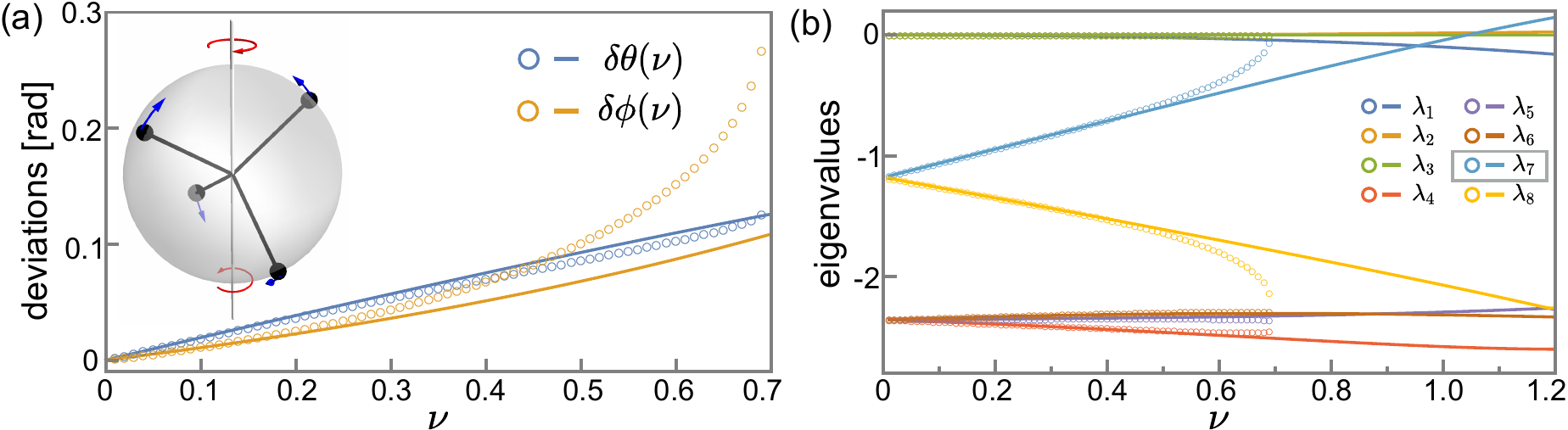}
	\caption{(a) Deviations of defect positions from the tetrahedron in the small activity regime. Shown are deviations $\delta\theta$ and $\delta\phi$ measured in simulations (circles) and obtained from an analytical solution for the stationary point at linear order (lines). {\it Inset:} the corresponding twist (red arrows) and stretching (blue arrows) modes of deformation of the initial tetrahedron. (b) Spectrum of $\grad \vect{g}$ evaluated at the ``skewed tetrahedron", using defect positions from the simulation (circles) and positions obtained from analytical expressions for $\delta\theta(\nu)$ and $\delta\phi(\nu)$ given in  \eqref{eq:ch4:solution_deltatheta} and \eqref{eq:ch4:solution_deltaphi} (lines). The data suggests that at $\nu^{*}=0.7$ the eigenvalue $\lambda_7$ (boxed) becomes positive, which renders the ``skewed tetrahedron" linearly unstable. This is qualitatively confirmed by the theoretical prediction, albeit with an overestimated transition point. }
	\label{fig:initialdeform}
\end{figure*}

For intermediate activity the positions of the four defects periodically pass through tetrahedral and planar configurations, as shown in Fig.\ \ref{fig:fourdefects} (a-c), which is the dynamics found in experiments~\cite{Keber:2014fh}. The motion is characterised by the formation of two counterrotating flow vortices that separate the defects into two pairs, in which they rotate around each other. This effect becomes more pronounced as the activity is increased, as shown in Fig.\ \ref{fig:fourdefects} (d-f). The separation of defects within each pair decreases significantly with $\nu$. There is also a gradual change in the shape of the trajectories, from square-like to more ellipsoid, such that the tetrahedral configurations are no longer approached and the defects oscillate between two different planar arrangements. As the defects in each pair are drawn closer with increasing activity the dynamics approaches the situation for two antipodal spirals in the director, which in the limit generate a perfectly symmetric flow vortex pair. This behaviour is summarised in Figure \ref{fig:parameterscan}, where the mean and the minimal angular distances are plotted against $\nu$, the latter reflecting the decreasing separation between defects in each pair. 

The total speed of the $+1/2$ defects, which also includes motion due to elasticity, is dominated by their active speed $v_0$ given by \eqref{eq:activespeed}. The frequency of the defect oscillations is thus 
\begin{equation}
f \sim \frac{v_0}{R} = \frac{h_0^2 \sigma_0}{\mu r_c R},
\end{equation}
without accounting for the small changes in the orbit shape with increasing $\nu$. 

The effective attraction of defects into pairs is mediated by the active flow vortices that form inbetween them, which in turn are controlled by the underlying director. In the tetrahedral configuration the nematic has a characteristic tennis ball texture \cite{Nelson:2002dd,Vitelli:2006ba}, but for generic values of $\alpha_0$ this texture is skewed, such that each two defects form a separated spiral. The flow vortices accordingly acquire a sink- or source-like component, depending on the tilt in the spiral. As can be seen for instance in Figure \ref{fig:fourdefects} (b), the paired defects have a sink-like vortex inbetween them which keeps them together. This active attraction mechanism requires the possibility of radial flows to accomodate this influx, which is guaranteed in the thin film approach. 

The choices $\alpha_0=0, \pi/2$ produce zero tilt in the texture of the initial tetrahedron and the resulting dynamics lacks the contraction of defect trajectories in one of the directions, such that they continue passing through tetrahedra for high activity. Finally, $\alpha_0=\pi/4$ does not have a dynamical regime and defects relax into increasingly tight, but stationary pairs. 


\begin{figure*}[th]
	\centering
	\includegraphics[width=0.99\textwidth]{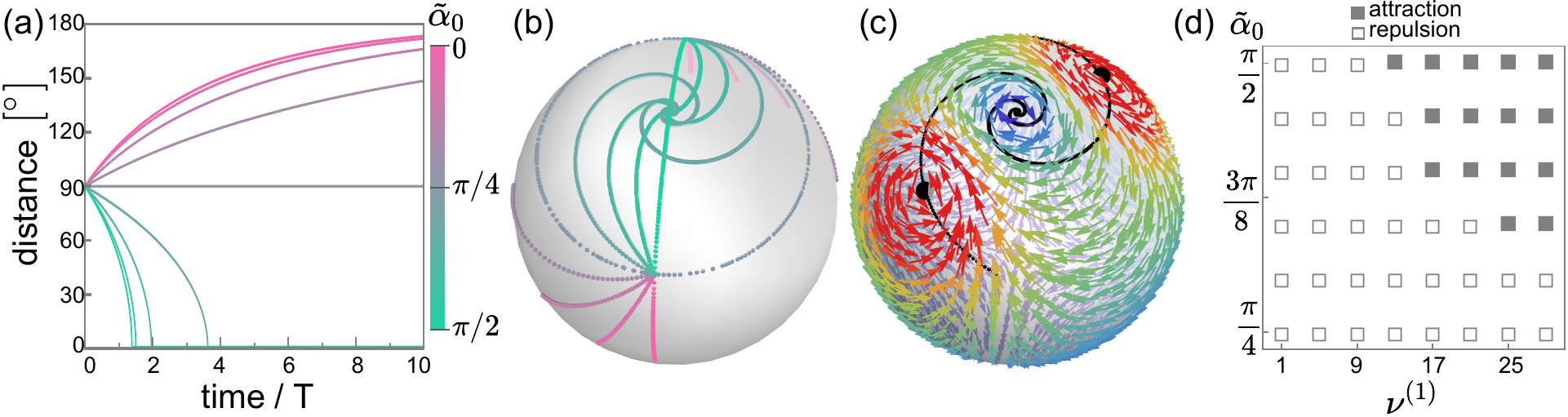}
	\caption{Active advection leads to repulsion of two aster-like defects and attraction of two vortex-like defects on a sphere. (a) Distance of the two +1 defects over time for different local director geometry, controlled by $\tilde{\alpha}_0$, which varies in steps of $\pi/16$. Here, the motion of defects is due to active advection only, with $\tilde{K}=0$. (b) Perfect asters ($\tilde{\alpha}_0=0$) or vortex defects ($\tilde{\alpha}_0=\pi/2$) move along geodesics, but in general the defects move on outward or inward spiralling trajectories. Two perfect spirals ($\tilde{\alpha}_0=\pi/4$) move along a circular path, without changing their distance. (c) Two spiral-like defects ($\tilde{\alpha}_0=5 \pi/16$) are attracted to each other by the flow vortex that forms inbetween them. (d) When elasticity is included, attraction of +1 defects is found only for $\nu^{(1)}$ above a threshold and for large enough tilt $\tilde{\alpha}_0$. \label{fig:twodefects_coalescing}}
\end{figure*}

\subsection{Linear stability of static configuration}
The defects only move above a critical $\nu^{*}\approx 0.7$ (Fig.\ \ref{fig:parameterscan}) and we describe this transition in a linear stability analysis. The system is initialised with the four defects at the vertices of a tetrahedron with $\theta_i^{(0)} = (\beta, \pi-\beta, \beta, \pi-\beta)$ and $ \phi_i ^{(0)}= (0,\pi/2, \pi, 3\pi/2)$, with $\beta =\arctan(\sqrt{2})$. It is evident from simulations that for activities below the threshold the defects settle into an increasingly distorted tetrahedron, which can be described by the coordinates
\begin{flalign}\label{eq:ch4:deviations1}
\theta_1^{*} &= \theta_3^{*} = \theta_1^{(0)} -\delta\theta,\\\label{eq:ch4:deviations2}
\theta_2^{*} &= \theta_4^{*} = \theta_2^{(0)} +\delta\theta,\\\label{eq:ch4:deviations3}
\phi_1^{*} &= \phi_1^{(0)} -\delta\phi, \,\, \phi_2^{*} = \phi_2^{(0)} +\delta\phi, \\\label{eq:ch4:deviations4}
\phi_3^{*} &= \phi_3^{(0)} -\delta\phi, \,\, \phi_4^{*} = \phi_4^{(0)} +\delta\phi,
\end{flalign}	
with small deviations $\delta \theta$ and $\delta \phi$ as shown in Fig.\ \ref{fig:initialdeform} (a). Using this ansatz we find analytical solutions for the deviations at linear order (see Appendix \ref{app:stability}). The deformation of the tetrahedron is a superposition of two modes -- twisting around and  stretching along the $z$-axis, illustrated in the inset of Fig.\ \ref{fig:initialdeform} (a).

At the critical activity the skewed tetrahedron becomes linearly unstable, as seen from the spectrum of the dynamical matrix $\grad \vect{g}$ (see Appendix \ref{app:stability} for definition) shown in Figure \ref{fig:initialdeform} (b). The simulation data suggests that one eigenvalue changes sign at $\nu^{*}$, while all others stay non-positive, indicating that the skewed tetrahedron is stable below $\nu^{*}$. The three vanishing eigenvalues correspond to rigid body rotations. Calculating the eigensystem using the analytical solutions for the deviations above the threshold allows to characterise this instability, albeit with an overestimated transition activity. The eigenvalue $\lambda_7$ becomes positive at $\nu\approx 1.0$, which marks the linear instability of the skewed tetrahedron towards a deformation that strongly increases the twist and slightly reverses the stretching. This can be seen from the corresponding eigenvector, which is of the form $(a,-a,a,-a,-b, b,-b,b)$ with $b\gg a>0$, and is exactly the dynamics found in simulations at the beginning of the periodic orbits (see Fig.\ \ref{fig:parameterscan} (b) and (c)).

\subsection{Two +1 defects}
ALCs can develop unit strength defects in their orientation \cite{1997Natur.389..305N,Kruse:2004il} and the confinement to a spherical shell provides a setup where two such defects could be topologically stabilised. We study the motion of two +1 defects in the limit of very strong activity, setting $\tilde{K}=0$ and using the active time scale $T=R^2\mu/h_0^2\sigma_0$. The value of $\alpha_0$ that is required to generate a particular director geometry at the defects depends on their position. This ambiguity can be removed by setting $\alpha_0 = - \arg(z_1-z_2) +\tilde{\alpha}_0$, where the additional constant is found by imposing an aster-geometry for $\tilde{\alpha}_0=0$ for both defects irrespective of their position. Now,  $\tilde{\alpha}_0\in[0,\pi/2]$ produces increasingly tilted spirals, up to two pure vortex defects for $\tilde{\alpha}_0=\pi/2$.

We find that two +1 defects are either attracted to or repelled from each other by active advection, depending on the local director geometry, as shown in Fig.\ \ref{fig:twodefects_coalescing} (a). Two defects that are aster-like ($0\leq \tilde{\alpha}_0 <\pi/4$) experience repulsion and relax into an antipodal configuration. Since the asters create sinks in the flow, a source-like flow vortex forms in between them, pushing them apart. Vortex-like defects ($\pi/4< \tilde{\alpha}_0 \leq \pi/2$) show the converse effect and are drawn towards each other. In this idealised setting without elasticity, they merge into a +2 boojum, with a local flow structure of a +3 singularity accompanied by a stagnation point at the antipodal point. Two perfect spiral defects ($\tilde{\alpha}_0=\pi/4$) keep a constant distance, rotating around each other on a circular path. 

During this motion the defects are typically spiralling inward or outward, as shown in Fig.\ \ref{fig:twodefects_coalescing} (b). Only in the two limiting cases do the defects move along their connecting geodesic. Figure \ref{fig:twodefects_coalescing} (c) shows the active flow with the additional sink-like vortex inbetween the defects ($\tilde{\alpha}_0=5 \pi/16$), that draws the defects inward on a spiralling trajectory. The defects' trajectory rotates in a direction opposite to the rotation of their local flow vortices. 

When elastic repulsion is included, with $\tilde{K}=1$, the defects relax into the antipodal configuration for all $\tilde{\alpha}_0$ for activities up to $\nu^{(1)}\approx 12$. Above this threshold active attraction overbalances the elastic repulsion for large enough $\tilde{\alpha}_0$, as shown in Fig.\ \ref{fig:twodefects_coalescing} (d). Interestingly, in such cases the defects again collapse into a very tight pair. In an experimental system,  fluctuations in the tilt of a spiral around the limiting value of $\tilde{\alpha}_0$ might lead to oscillations between the antipodal and the collapsed configurations.   

\begin{figure*}[t]
	\includegraphics[width=0.99\linewidth]{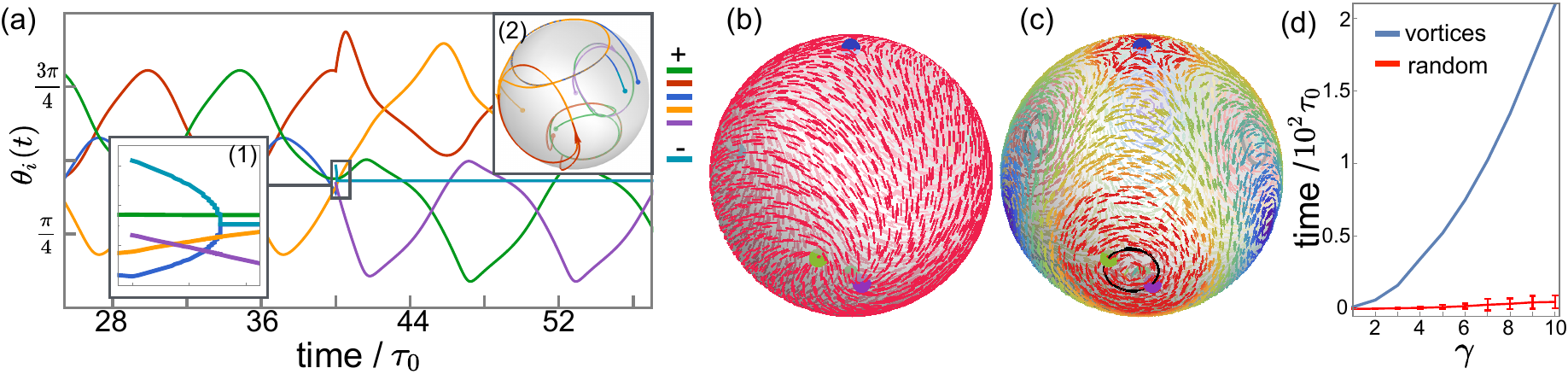}
	\caption{(a) Four $+1/2$ defects show regular oscillations before and shortly after the insertion of an additional $\pm 1/2$ defect pair, as seen from $\theta_i(t)$ for $i=1,\dots,6$ with $\gamma=2$. {\it Inset 1:} Close-up on the rapid annihilation of the additional pair. {\it Inset 2:} Defects resume similar trajectories after the fluctuation, but with different pairing. (b) Orientation field with eight defects, six +1/2 and two $-1/2$, producing a metastable flow vortex arrangement. (c) Corresponding flow structure with six equidistant vortices on the equator, +1/2 defect trajectories marked in black. (d) Time to first annihilation event over $\gamma$, determining the shell radius as $R=\gamma R_0$, for the vortex configuration in (b,c) and averaged over 500 random initial defect positions. \label{fig:manydefects}}
\end{figure*}

\subsection{Many-defect states}
When the activity $|\sigma_0|$ or the shell size $R$ are increased additional $\pm 1/2$ defect pairs may be created on top of the four +1/2 defects, as the system approaches the onset of active turbulence \cite{2015PhRvX...5c1003G}. To study such situations we increase the radius as $R=\gamma R_0$, with $\gamma>1$, keeping all other parameters fixed. This changes the elastic time scale to $\tau = \gamma^2 \tau_0$ and the activity-to-elasticity ratio to $\nu = \gamma \nu_0$. The reference values correspond to parameters in Section III. A for the regime of tetrahedral-planar oscillations, for instance $\nu_0=1$. 

We consider a system with four defects in this oscillatory state and inject one $\pm1/2$ pair at a random position. Figure \ref{fig:manydefects} (a) shows how the dynamics reacts to this perturbation. One of the $+1/2$ defects very quickly annihilates with the $-1/2$ and the remaining four defects resume the oscillation, usually in a different pairing. Similarly, when all defects are placed at random positions the annihilation events happen rapidly, leaving the minimal four-defect state in the oscillating regime. The same is found for more than one additional pair of half-defects in the system. This indicates that the oscillatory state is stable, as long as additional defect pairs occur as fluctuations and are not produced constantly.

On the other hand, by inducing additional $\pm1/2$ defects at specific locations more complex flow vortex configurations may be constructed, in which elastic forces and active flows are balanced. The simplest many-defect configuration that is metastable has $n_{\mathrm{pair}}=2$ additional defect pairs, shown in Fig.\ \ref{fig:manydefects} (b,c). The six +1/2 defects are allocated to three flow vortices arranged equidistantly around the equator, with another three vortices rotating in the opposite direction inbetween them. The three-fold symmetry of this flow field is guided by the flow singularities at the $-1/2$ defects, which have $\mathcal{I}=-2$ and are located at the poles. This configuration is transient and reduces to the four-defect state due to coalescence of oppositely charged defects. In Figure \ref{fig:manydefects} (d) the times to the first annihilation event for this metastable configuration and for $n_\mathrm{pair}=2$ with random initial defect positions are compared. For the vortex configuration this time increases considerably with $\gamma$. This opens an interesting direction of tuning specific many-defect states before the onset of active turbulence by exploiting the topologically required singularities in  both flow and director. The metastability of such configurations could be aided by an advantageous manipulation of the shell shape, for instance by trapping positive defects in regions of higher curvature \cite{Selinger:2011ky}.


\section{Discussion}
In contact with a passive fluid on the outside the active shell may swim due to its self-generated surface flows. In a squirmer approach the swimming velocity of a spherical shell can be calculated as the surface integral of the slip velocity, $\vect{U}(t) = - \frac{1}{4\pi R^2} \int_{S} \vect{u}_{\perp}(t) \mathrm{d}S$, with a similar expression for the angular velocity $\vect{\Omega}(t)$ \cite{Stone:1996ir}. Evaluating the integrals exactly, we find that a single +2 defect as well as two +1 defects do not generate translational or rotational motion of the shell, irrespective of their position. The same results from numerical integration of the slip velocity for four half-defects along their symmetric trajectories. There is a number of ways in which this symmetry could be broken in order to achieve self-propulsion and rotation, including distortions of the vortices by noise, changes in vesicle shape \cite{10.1039/9781782628491-00313}, and interactions with surfaces or other active vesicles \cite{Sanchez:2013gt}. One promising direction is a controlled asymmetric modification of the local director geometry around two +1 defects \cite{futurepaper}.  

Our results can be extended to contractile active fluids by changing the sign of the activity $\sigma_0$. The reversed sign of the flow exchanges the role of splay-like and bend-like distortions in the orientation. The direction of motion of half-integer defects is reversed, but the tetrahedral-planar oscillations and the formation of vortices -- with opposite rotation sense -- is unchanged. Similarly, the type of active interaction between +1 defects is reversed. The thin film approach used here allows for a non-zero radial flow component, which in general is present in the examples considered and enables the repulsion or attraction of defects due to active flows. The radial component is small, $\sim \mathcal{O}(h_0/R)$, compared to tangential flows and will result in a dynamic deviation from the spherical shape that complements the defect motion \cite{futurepaper}. Stationary shell shapes, e.g. for two asters, should locally resemble profiles found for flat droplets of active nematics with defects \cite{Joanny:2012dx, Khoromskaia:2015ec}. We have taken a one elastic constant approximation for simplicity, but in systems of elongated filaments one can expect anisotropy. If this is sufficiently large it could lead to qualitative changes of the dynamics and is certainly an extension worth pursuing. Our work establishes the formation of vortices under confinement as a generic feature also for ALCs on spherical surfaces. It would be interesting to extend to other topologies, for instance tori with additional handles \cite{Pairam04062013}. 

We thank Carl Whitfield, Daniel Pearce and Julia Yeomans for enligtening discussions on various aspects of this work. This work was supported by the UK EPSRC through Grant No. A.MACX.0002. 

\section{Appendix}

\subsection{Flow in a thin active nematic shell}\label{app:flow}
{\footnotesize
	In a thin spherical film of active nematic the generalised Stokes equation given in the main text can be expanded in the small parameter $\varepsilon = h_0/R$. Using the typical magnitude  $U_0=R/T$ of active flows, with the time scale $T$, the dimensionless velocity components are $\tilde{u}_{\theta}=u_{\theta}/U_0$, $\tilde{u}_{\phi}=u_{\phi}/U_0$, and $\tilde{u}_r= u_r/\varepsilon U_0$. The radial coordinate becomes 
	\[\tilde{r}= \frac{1}{\varepsilon}\left(\frac{r}{R}-1\right)\,, \]
	and the corresponding partial derivatives $\partial_r f= \frac{1}{h_0}\partial_{\tilde{r}}f$, for some function $f(r)$. In the dimensionless form, for instance the $\theta$-component of the Stokes equation becomes
	\begin{equation}\label{app:stokesdimless}
	0 = - \frac{\varepsilon^2 R}{\mu U_0} \dtheta p + \frac{\varepsilon^2 R^2}{\mu U_0}  \left(\nabla \cdot \vect{\sigma}^a \right)_{\theta} + \partial_{\tilde{r}_s}^2\tilde{u}_{\theta}+\mathcal{O}(\varepsilon^2)\,,
	\end{equation}
	and there is a similar expression for the $\phi$-component. The active stress tensor gradient is 
	\begin{equation}
	\nabla \cdot \vect{\sigma}^a  = - \frac{\sigma_0}{R} \left( 
	\begin{array}{c}
	1\\
	- \sin\left( 2 \psi \right) \dtheta \psi + \frac{ \cos\left(2 \psi\right)}{\sin \theta}\left( \cos \theta +  \dphi \psi \right)\\
	\cos \left( 2 \psi \right) \dtheta \psi +  \frac{\sin (2 \psi)}{\sin\theta}\left(\cos \theta +\dphi \psi  \right) 
	\end{array}
	\right).
	\end{equation}
	Therefore, in order for the activity to drive the tangential flows the corresponding prefactor in \eqref{app:stokesdimless} has to scale as $\sim\mathcal{O}(1)$, leading to the dimensionless prefactor
	\begin{equation}\label{eq:activescaling}
	\tilde{\sigma}_0 =  \frac{\varepsilon^2 R}{\mu U_0}\sigma_0\,
	\end{equation}
	and a similar relation for the pressure. Analogous to planar thin films \cite{1997RvMP...69..931O,Khoromskaia:2015ec}, the leading order part of the $r$-component of the Stokes equation yields a constant pressure. With the boundary conditions 
	\begin{equation}
	\partial_{\tilde{r}_s} \tilde{u}_{\phi}|_{\tilde{r}_s=1}=0 \quad \mathrm{and} \quad \tilde{u}_{\phi}|_{\tilde{r}_s=0}=0\,,
	\end{equation}   
	the solution \eqref{eq:flow} for the tangential flow components  is obtained. 
	
	For a fixed $\tilde{r}$, for instance $\tilde{r}=1$ used for the defect dynamics, the tangential flow can be written in a complex representation making use of the stereographic projection $z(\theta,\phi) = R \cot(\theta/2)e^{i\phi}$. In this way, the projection point is the north pole and the plane crosses the sphere along the equator. The complex flow is given by
	\begin{align}\nonumber
	\tilde{u}(z,\bar{z})&= \tilde{u}_{\theta}+i\tilde{u}_{\phi}\\ \label{eq:complexflow}
	&= -\frac{\tilde{\sigma}_0}{2} e^{-i2\alpha(z,\bar{z})}\bigg\{\frac{z^2}{R\abs{z}} -\frac{z}{2}\bigg(\frac{R}{\abs{z}}+ \frac{\abs{z}}{R}\bigg) \sum_{j=1}^{n_{\mathrm{def}}} m_j \frac{z-z_j}{\abs{z-z_j}^2}\bigg\} 
	\end{align}
	This expression is well-defined through the stereographic projection of the tangential flow onto the plane, which is given in the complex form as $u=u_x+iu_y$ and relates to $\tilde{u}$ as
	\begin{equation}
	\tilde{u} = -(1-\cos\theta) e^{i\phi} \bar{u}\,.
	\end{equation}
	We evaluate \eqref{eq:complexflow} on the projection of the small circle $\gamma_k(s)$ introduced in the main text, which has the form
	\begin{equation}
	z(s) = z_k - \frac{\rho R}{1 -\cos\theta_k}e^{i(\phi_k -s)}\
	\end{equation}
	with $s\in[0,2\pi]$, provided the circle does not enclose one of the poles on the sphere. An expansion of \eqref{eq:complexflow} in powers of $\rho$ reads
	\begin{align}
	\tilde{u}(\rho)\nonumber
	&= \frac{m_k}{\rho} e^{i(2m_k -1)s} e^{i2(1-m_k)\phi_k} e^{-i2w(z_k)} \\ \nonumber
	& + e^{-i2w(z_k)} e^{-i 2m_k\phi_k}\bigg\{ \frac{m_k}{2}e^{i2(m_k-1)s}e^{i2\phi_k} h_1(z_1,\bar{z}_1,\dots)\\ \label{app:expansion} &+  e^{i2m_k s} h_2(z_1,\bar{z}_1,\dots) \bigg\}
	+ \mathcal{O}(\rho)
	\end{align}
	where the functions $h_1$ and $h_2$ only depend on the defect positions and other constants. Integrating this expression over $s$ yields a non-winding contribution at the order $\mathcal{O}(1/\rho)$ for $m_k=1/2$ and at the order $\mathcal{O}(1)$ for $m_k=1$, as discussed in the main text. 
}

\subsection{Point-particle-like dynamics of defects}\label{app:dynamics}

{\footnotesize 
	The free energy of a nematic on a sphere can be phrased in terms of the defects' pairwise interaction energies and self-energies \cite{Lubensky:1992bn,Nelson:2002dd,Vitelli:2006ba}, which are constant in our model, 
	\begin{equation}\label{eq:ch4:energydefects}
	E = -\frac{\pi K}{2} \sum_{\substack{i,j =1 \\ i \neq j }}^{n_\mathrm{def}} m_i m_j \ln(1-\cos \beta_{ij}) + const.
	\end{equation} 
	The angular distance between defect $i$ and $j$ is given by  
	\begin{equation}\label{eq:ch4:geodesic}
	\cos \beta_{ij} = \cos\theta_i \cos\theta_j + \sin\theta_i \sin\theta_j \cos(\phi_i-\phi_j)\,.
	\end{equation}
	The force acting on defect $k$ due to all other defects is \cite{Chaikin,Keber:2014fh}
	\begin{equation}\label{eq:ch4:elasticforce}
	\vect{F}^{(k)} = - \grad_k E = -\bigg(\hat{\vect{e}}_{\theta,k}\frac{1}{R} \partial_{\theta_k}E + \hat{\vect{e}}_{\phi,k} \frac{1}{R \sin\theta_k}\partial_{\phi_k} E\bigg)\,,
	\end{equation}
	where the notation $\hat{\vect{e}}_{\theta,k}= \etheta(\theta_k,\phi_k)$ and $\hat{\vect{e}}_{\phi,k}= \ephi(\theta_k,\phi_k)$ is used. The $\theta$-component of \eqref{eq:ch4:elasticforce} contains
	\begin{equation}
	\partial_{\theta_k} E = K\pi m_k \sum_{j=1,\,  j\neq k}^{n_\mathrm{def}} m_j \frac{\partial_{\theta_k}\cos\beta_{kj} }{1-\cos\beta_{kj}}  
	\end{equation}
	and the expression for $\partial_{\phi_k} E$ is analogous. The elastic terms in the dimensionless dynamical equations \eqref{dynamicsnondim1} and \eqref{dynamicsnondim2} can be written as
	\begin{equation}
	\frac{\tau}{\xi R} F_{\theta}^{(k)} = - \frac{\tau K}{\xi R^2} \pi m_k \sum_{j=1,\,  j\neq k}^{n_{\mathrm{def}}} m_j \frac{\partial_{\theta_k}\cos\beta_{kj} }{1-\cos\beta_{kj}}\,,
	\end{equation}
	and a similar expression for the $\phi$-component. Making time dimensionless with $\tau$ leads to $\tilde{K}=\tau K/\xi R^2=1$.
	
	For example, for four $+1/2$ defects the full dynamical system reads
	\begin{flalign}\label{eq:ch4:dynamicsfour1}
	\partial_{\tilde{t}}\theta_k &= -\frac{\pi}{4} \sum_{j=1,\,  j\neq k}^{4} \frac{\partial_{\theta_k}\cos\beta_{kj} }{1-\cos\beta_{kj}} - \frac{\nu}{4} \cos\left( \phi_k - 2w(z_k)  \right)\\ \label{eq:ch4:dynamicsfour2}
	\partial_{\tilde{t}}\phi_k &= \frac{1}{\sin\theta_k} \left( - \frac{\pi}{4\sin\theta_k} \sum_{j=1,\,  j\neq k}^{4} \frac{\partial_{\phi_k}\cos\beta_{kj} }{1-\cos\beta_{kj}} - \frac{\nu}{4} \sin\left( \phi_k - 2w(z_k)  \right)\right)
	\end{flalign}
	for $k=1, \dots,4 $, where $w(z_k) = \alpha_0 +\frac{\pi}{2} + \frac{1}{2}\sum_{j \neq k} \Im\{ \ln(z_k-z_j )\}$ and $\nu$ is defined in \eqref{eq:ratio}. 
	
	For all defect configurations the dynamical systems are integrated using the ordinary differential equation solver {\tt ode23s} provided by the software {\it MATLAB 2016a}, with relative and absolute accuracies set to $10^{-6}\tau$.
}

\subsection{Analytical solutions for a small deviation from the tetrahedron}\label{app:stability}
{\footnotesize
	In order to study the linear stability of the skewed tetrahedron the four defects settle into for low activity we write their dynamical equations as
	\begin{equation}\label{eq:ch4:dynamicsfour3}
	\dv{\vect{x}(t)}{t} = \vect{g}(\vect{x}(t))\,,
	\end{equation} 
	where $\vect{x}(t)=\left(\theta_1(t),\dots,\theta_4(t),\phi_1(t),\dots,\phi_4(t)\right) \in \mathbb{R}^8$ is the vector of spherical defect coordinates and $\vect{g}(\vect{x}(t))$ are the concatenated right hand sides of equations \eqref{eq:ch4:dynamicsfour1} and \eqref{eq:ch4:dynamicsfour2}. Motivated by the defect motion in simulations we use the ansatz given by \eqref{eq:ch4:deviations1} -- \eqref{eq:ch4:deviations4} for the fixed point $\vect{x}^{*}$ representing the skewed tetrahedron, with small deviations $0<\delta\theta, \delta\phi \ll 1$. With this ansatz, the stationary condition
	\begin{equation}
	\vect{g}(\vect{x}(\delta\theta,\delta\phi)) = \vect{0}\,,
	\end{equation}	
	linearised in $\delta \theta$ and $\delta \phi$, has the solutions
	\begin{flalign}\label{eq:ch4:solution_deltatheta}
	\delta\theta (\nu,\alpha_0) & = -\frac{2 \nu  \left(\pi  \cos (2 \alpha_0)+\sqrt{2} \nu \right)}{2 \nu ^2 \sin ^2(2 \alpha_0)+3 \pi  \sqrt{2} \nu  \cos (2 \alpha_0)+3 \pi ^2},\\ \label{eq:ch4:solution_deltaphi}
	\delta\phi(\nu,\alpha_0) &= \frac{2 \sqrt{3} \nu  \left(3 \sqrt{2} \pi -2 \nu  \cos (2 \alpha_0)\right)}{12 \nu ^2 \sin (2 \alpha_0)+9 \pi  \csc (\alpha_0) \sec (\alpha_0) \left(\sqrt{2} \nu  \cos (2 \alpha_0)+\pi \right)}, 
	\end{flalign}	
	which are plotted in Figure \ref{fig:initialdeform} (a) together with the the deviations measured in the simulations. A perturbation $\delta \vect{x}$ away from $\vect{x}^{*}$ evolves according to 
		\begin{equation}
	\dv{\delta \vect{x}}{t} = \grad \vect{g} \big|_{\vect{x}^{*}}\cdot \delta \vect{x},
	\end{equation} 
	and the spectrum of the dynamical matrix  
	\begin{equation}\label{eq:ch4:eigenvalues}
	\grad \vect{g} \big|_{\vect{x}^{*}= (\theta_1^{*},\dots, \phi_4^{*})}\,,
	\end{equation}
	characterises the linear stability of $\vect{x}^{*}$. The eigenvalues are plotted in Figure \ref{fig:initialdeform} (b), calculated numerically from measured deviations and from the linear solution \eqref{eq:ch4:solution_deltatheta} and \eqref{eq:ch4:solution_deltaphi}, respectively. 
	
}

\bibliography{activeshell} 

\begin{thebibliography}{49}%
\makeatletter
\providecommand \@ifxundefined [1]{%
 \@ifx{#1\undefined}
}%
\providecommand \@ifnum [1]{%
 \ifnum #1\expandafter \@firstoftwo
 \else \expandafter \@secondoftwo
 \fi
}%
\providecommand \@ifx [1]{%
 \ifx #1\expandafter \@firstoftwo
 \else \expandafter \@secondoftwo
 \fi
}%
\providecommand \natexlab [1]{#1}%
\providecommand \enquote  [1]{``#1''}%
\providecommand \bibnamefont  [1]{#1}%
\providecommand \bibfnamefont [1]{#1}%
\providecommand \citenamefont [1]{#1}%
\providecommand \href@noop [0]{\@secondoftwo}%
\providecommand \href [0]{\begingroup \@sanitize@url \@href}%
\providecommand \@href[1]{\@@startlink{#1}\@@href}%
\providecommand \@@href[1]{\endgroup#1\@@endlink}%
\providecommand \@sanitize@url [0]{\catcode `\\12\catcode `\$12\catcode
  `\&12\catcode `\#12\catcode `\^12\catcode `\_12\catcode `\%12\relax}%
\providecommand \@@startlink[1]{}%
\providecommand \@@endlink[0]{}%
\providecommand \url  [0]{\begingroup\@sanitize@url \@url }%
\providecommand \@url [1]{\endgroup\@href {#1}{\urlprefix }}%
\providecommand \urlprefix  [0]{URL }%
\providecommand \Eprint [0]{\href }%
\providecommand \doibase [0]{http://dx.doi.org/}%
\providecommand \selectlanguage [0]{\@gobble}%
\providecommand \bibinfo  [0]{\@secondoftwo}%
\providecommand \bibfield  [0]{\@secondoftwo}%
\providecommand \translation [1]{[#1]}%
\providecommand \BibitemOpen [0]{}%
\providecommand \bibitemStop [0]{}%
\providecommand \bibitemNoStop [0]{.\EOS\space}%
\providecommand \EOS [0]{\spacefactor3000\relax}%
\providecommand \BibitemShut  [1]{\csname bibitem#1\endcsname}%
\let\auto@bib@innerbib\@empty
\bibitem [{\citenamefont {Prost}\ \emph {et~al.}(2015)\citenamefont {Prost},
  \citenamefont {J{\"u}licher},\ and\ \citenamefont
  {Joanny}}]{2015NatPh..11..111P}%
  \BibitemOpen
  \bibfield  {author} {\bibinfo {author} {\bibfnamefont {J.}~\bibnamefont
  {Prost}}, \bibinfo {author} {\bibfnamefont {F.}~\bibnamefont {J{\"u}licher}},
  \ and\ \bibinfo {author} {\bibfnamefont {J.~F.}\ \bibnamefont {Joanny}},\
  }\bibfield  {title} {\enquote {\bibinfo {title} {Active gel physics},}\
  }\href {\doibase 10.1038/nphys3224} {\bibfield  {journal} {\bibinfo
  {journal} {Nat. Phys.}\ }\textbf {\bibinfo {volume} {11}},\ \bibinfo {pages}
  {111} (\bibinfo {year} {2015})}\BibitemShut {NoStop}%
\bibitem [{\citenamefont {Marchetti}\ \emph {et~al.}(2013)\citenamefont
  {Marchetti}, \citenamefont {Joanny}, \citenamefont {Ramaswamy}, \citenamefont
  {Liverpool}, \citenamefont {Prost}, \citenamefont {Rao},\ and\ \citenamefont
  {Simha}}]{Marchetti:2013bp}%
  \BibitemOpen
  \bibfield  {author} {\bibinfo {author} {\bibfnamefont {M.~C.}\ \bibnamefont
  {Marchetti}}, \bibinfo {author} {\bibfnamefont {J.~F.}\ \bibnamefont
  {Joanny}}, \bibinfo {author} {\bibfnamefont {S.}~\bibnamefont {Ramaswamy}},
  \bibinfo {author} {\bibfnamefont {T.~B.}\ \bibnamefont {Liverpool}}, \bibinfo
  {author} {\bibfnamefont {J.}~\bibnamefont {Prost}}, \bibinfo {author}
  {\bibfnamefont {M.}~\bibnamefont {Rao}}, \ and\ \bibinfo {author}
  {\bibfnamefont {R.~A.}\ \bibnamefont {Simha}},\ }\bibfield  {title} {\enquote
  {\bibinfo {title} {Hydrodynamics of soft active matter},}\ }\href {\doibase
  10.1103/RevModPhys.85.1143} {\bibfield  {journal} {\bibinfo  {journal} {Rev.
  Mod. Phys.}\ }\textbf {\bibinfo {volume} {85}},\ \bibinfo {pages} {1143}
  (\bibinfo {year} {2013})}\BibitemShut {NoStop}%
\bibitem [{\citenamefont {Ramaswamy}(2010)}]{Ramaswamy:2010bf}%
  \BibitemOpen
  \bibfield  {author} {\bibinfo {author} {\bibfnamefont {S.}~\bibnamefont
  {Ramaswamy}},\ }\bibfield  {title} {\enquote {\bibinfo {title} {The mechanics
  and statistics of active matter},}\ }\href {\doibase
  10.1146/annurev-conmatphys-070909-104101} {\bibfield  {journal} {\bibinfo
  {journal} {Annu. Rev. Condens. Matter Phys.}\ }\textbf {\bibinfo {volume}
  {1}},\ \bibinfo {pages} {323} (\bibinfo {year} {2010})}\BibitemShut {NoStop}%
\bibitem [{\citenamefont {Kruse}\ \emph {et~al.}(2006)\citenamefont {Kruse},
  \citenamefont {Joanny}, \citenamefont {J{\"u}licher},\ and\ \citenamefont
  {Prost}}]{Kruse:2006gq}%
  \BibitemOpen
  \bibfield  {author} {\bibinfo {author} {\bibfnamefont {K.}~\bibnamefont
  {Kruse}}, \bibinfo {author} {\bibfnamefont {J.~F.}\ \bibnamefont {Joanny}},
  \bibinfo {author} {\bibfnamefont {F.}~\bibnamefont {J{\"u}licher}}, \ and\
  \bibinfo {author} {\bibfnamefont {J.}~\bibnamefont {Prost}},\ }\bibfield
  {title} {\enquote {\bibinfo {title} {Contractility and retrograde flow in
  lamellipodium motion},}\ }\href {\doibase 10.1088/1478-3975/3/2/005}
  {\bibfield  {journal} {\bibinfo  {journal} {Phys. Biol.}\ }\textbf {\bibinfo
  {volume} {3}},\ \bibinfo {pages} {130} (\bibinfo {year} {2006})}\BibitemShut
  {NoStop}%
\bibitem [{\citenamefont {Tjhung}\ \emph {et~al.}(2015)\citenamefont {Tjhung},
  \citenamefont {Tiribocchi}, \citenamefont {Marenduzzo},\ and\ \citenamefont
  {Cates}}]{2015NatCo...6E5420T}%
  \BibitemOpen
  \bibfield  {author} {\bibinfo {author} {\bibfnamefont {E.}~\bibnamefont
  {Tjhung}}, \bibinfo {author} {\bibfnamefont {A.}~\bibnamefont {Tiribocchi}},
  \bibinfo {author} {\bibfnamefont {D.}~\bibnamefont {Marenduzzo}}, \ and\
  \bibinfo {author} {\bibfnamefont {M.~E.}\ \bibnamefont {Cates}},\ }\bibfield
  {title} {\enquote {\bibinfo {title} {A minimal physical model captures the
  shapes of crawling cells},}\ }\href {\doibase 10.1038/ncomms6420} {\bibfield
  {journal} {\bibinfo  {journal} {Nat. Commun.}\ }\textbf {\bibinfo {volume}
  {6}},\ \bibinfo {pages} {5420} (\bibinfo {year} {2015})}\BibitemShut
  {NoStop}%
\bibitem [{\citenamefont {Tjhung}\ \emph {et~al.}(2012)\citenamefont {Tjhung},
  \citenamefont {Marenduzzo},\ and\ \citenamefont
  {Cates}}]{2012PNAS..10912381T}%
  \BibitemOpen
  \bibfield  {author} {\bibinfo {author} {\bibfnamefont {E.}~\bibnamefont
  {Tjhung}}, \bibinfo {author} {\bibfnamefont {D.}~\bibnamefont {Marenduzzo}},
  \ and\ \bibinfo {author} {\bibfnamefont {M.~E.}\ \bibnamefont {Cates}},\
  }\bibfield  {title} {\enquote {\bibinfo {title} {Spontaneous symmetry
  breaking in active droplets provides a generic route to motility},}\ }\href
  {\doibase 10.1073/pnas.1200843109} {\bibfield  {journal} {\bibinfo  {journal}
  {Proc. Natl. Acad. Sci. U.S.A.}\ }\textbf {\bibinfo {volume} {109}},\
  \bibinfo {pages} {12381} (\bibinfo {year} {2012})}\BibitemShut {NoStop}%
\bibitem [{\citenamefont {Salbreux}\ \emph {et~al.}(2009)\citenamefont
  {Salbreux}, \citenamefont {Prost},\ and\ \citenamefont
  {Joanny}}]{Salbreux:2009fp}%
  \BibitemOpen
  \bibfield  {author} {\bibinfo {author} {\bibfnamefont {G.}~\bibnamefont
  {Salbreux}}, \bibinfo {author} {\bibfnamefont {J.}~\bibnamefont {Prost}}, \
  and\ \bibinfo {author} {\bibfnamefont {J.~F.}\ \bibnamefont {Joanny}},\
  }\bibfield  {title} {\enquote {\bibinfo {title} {Hydrodynamics of cellular
  cortical flows and the formation of contractile rings},}\ }\href {\doibase
  10.1103/PhysRevLett.103.058102} {\bibfield  {journal} {\bibinfo  {journal}
  {Phys. Rev. Lett.}\ }\textbf {\bibinfo {volume} {103}},\ \bibinfo {pages}
  {058102} (\bibinfo {year} {2009})}\BibitemShut {NoStop}%
\bibitem [{\citenamefont {Turlier}\ \emph {et~al.}(2014)\citenamefont
  {Turlier}, \citenamefont {Audoly}, \citenamefont {Prost},\ and\ \citenamefont
  {Joanny}}]{Turlier:2014hq}%
  \BibitemOpen
  \bibfield  {author} {\bibinfo {author} {\bibfnamefont {H.}~\bibnamefont
  {Turlier}}, \bibinfo {author} {\bibfnamefont {B.}~\bibnamefont {Audoly}},
  \bibinfo {author} {\bibfnamefont {J.}~\bibnamefont {Prost}}, \ and\ \bibinfo
  {author} {\bibfnamefont {J.~F.}\ \bibnamefont {Joanny}},\ }\bibfield  {title}
  {\enquote {\bibinfo {title} {Furrow constriction in animal cell
  cytokinesis},}\ }\href {\doibase 10.1016/j.bpj.2013.11.014} {\bibfield
  {journal} {\bibinfo  {journal} {Biophys. J.}\ }\textbf {\bibinfo {volume}
  {106}},\ \bibinfo {pages} {114} (\bibinfo {year} {2014})}\BibitemShut
  {NoStop}%
\bibitem [{\citenamefont {Brugu{\'e}s}\ and\ \citenamefont
  {Needleman}(2014)}]{Brugues:2014bp}%
  \BibitemOpen
  \bibfield  {author} {\bibinfo {author} {\bibfnamefont {J.}~\bibnamefont
  {Brugu{\'e}s}}\ and\ \bibinfo {author} {\bibfnamefont {D.}~\bibnamefont
  {Needleman}},\ }\bibfield  {title} {\enquote {\bibinfo {title} {Physical
  basis of spindle self-organization},}\ }\href {\doibase
  10.1073/pnas.1409404111} {\bibfield  {journal} {\bibinfo  {journal} {Proc.
  Natl. Acad. Sci. U.S.A.}\ }\textbf {\bibinfo {volume} {111}},\ \bibinfo
  {pages} {18496} (\bibinfo {year} {2014})}\BibitemShut {NoStop}%
\bibitem [{\citenamefont {Salbreux}\ \emph {et~al.}(2007)\citenamefont
  {Salbreux}, \citenamefont {Joanny}, \citenamefont {Prost},\ and\
  \citenamefont {Pullarkat}}]{Salbreux:2007fc}%
  \BibitemOpen
  \bibfield  {author} {\bibinfo {author} {\bibfnamefont {G.}~\bibnamefont
  {Salbreux}}, \bibinfo {author} {\bibfnamefont {J.~F.}\ \bibnamefont
  {Joanny}}, \bibinfo {author} {\bibfnamefont {J.}~\bibnamefont {Prost}}, \
  and\ \bibinfo {author} {\bibfnamefont {P.}~\bibnamefont {Pullarkat}},\
  }\bibfield  {title} {\enquote {\bibinfo {title} {Shape oscillations of
  non-adhering fibroblast cells},}\ }\href {\doibase 10.1088/1478-3975/4/4/004}
  {\bibfield  {journal} {\bibinfo  {journal} {Phys. Biol.}\ }\textbf {\bibinfo
  {volume} {4}},\ \bibinfo {pages} {268} (\bibinfo {year} {2007})}\BibitemShut
  {NoStop}%
\bibitem [{\citenamefont {Callan-Jones}\ \emph {et~al.}(2016)\citenamefont
  {Callan-Jones}, \citenamefont {Ruprecht}, \citenamefont {Wieser},
  \citenamefont {Heisenberg},\ and\ \citenamefont
  {Voituriez}}]{CallanJones:2016fe}%
  \BibitemOpen
  \bibfield  {author} {\bibinfo {author} {\bibfnamefont {A.~C.}\ \bibnamefont
  {Callan-Jones}}, \bibinfo {author} {\bibfnamefont {V.}~\bibnamefont
  {Ruprecht}}, \bibinfo {author} {\bibfnamefont {S.}~\bibnamefont {Wieser}},
  \bibinfo {author} {\bibfnamefont {C.~P.}\ \bibnamefont {Heisenberg}}, \ and\
  \bibinfo {author} {\bibfnamefont {R.}~\bibnamefont {Voituriez}},\ }\bibfield
  {title} {\enquote {\bibinfo {title} {Cortical flow-driven shapes of
  nonadherent cells},}\ }\href {\doibase 10.1103/PhysRevLett.116.028102}
  {\bibfield  {journal} {\bibinfo  {journal} {Phys. Rev. Lett.}\ }\textbf
  {\bibinfo {volume} {116}},\ \bibinfo {pages} {028102} (\bibinfo {year}
  {2016})}\BibitemShut {NoStop}%
\bibitem [{\citenamefont {Doostmohammadi}\ \emph {et~al.}(2016)\citenamefont
  {Doostmohammadi}, \citenamefont {Thampi},\ and\ \citenamefont
  {Yeomans}}]{Doostmohammadi:2016iq}%
  \BibitemOpen
  \bibfield  {author} {\bibinfo {author} {\bibfnamefont {A.}~\bibnamefont
  {Doostmohammadi}}, \bibinfo {author} {\bibfnamefont {S.~P.}\ \bibnamefont
  {Thampi}}, \ and\ \bibinfo {author} {\bibfnamefont {J.~M.}\ \bibnamefont
  {Yeomans}},\ }\bibfield  {title} {\enquote {\bibinfo {title} {Defect-mediated
  morphologies in growing cell colonies},}\ }\href {\doibase
  10.1103/PhysRevLett.117.048102} {\bibfield  {journal} {\bibinfo  {journal}
  {Phys. Rev. Lett.}\ }\textbf {\bibinfo {volume} {117}},\ \bibinfo {pages}
  {048102} (\bibinfo {year} {2016})}\BibitemShut {NoStop}%
\bibitem [{\citenamefont {Wioland}\ \emph {et~al.}(2013)\citenamefont
  {Wioland}, \citenamefont {Woodhouse}, \citenamefont {Dunkel}, \citenamefont
  {Kessler},\ and\ \citenamefont {Goldstein}}]{Wioland:2013jm}%
  \BibitemOpen
  \bibfield  {author} {\bibinfo {author} {\bibfnamefont {H.}~\bibnamefont
  {Wioland}}, \bibinfo {author} {\bibfnamefont {F.~G.}\ \bibnamefont
  {Woodhouse}}, \bibinfo {author} {\bibfnamefont {J.}~\bibnamefont {Dunkel}},
  \bibinfo {author} {\bibfnamefont {J.~O.}\ \bibnamefont {Kessler}}, \ and\
  \bibinfo {author} {\bibfnamefont {R.~E.}\ \bibnamefont {Goldstein}},\
  }\bibfield  {title} {\enquote {\bibinfo {title} {Confinement stabilizes a
  bacterial suspension into a spiral vortex},}\ }\href {\doibase
  10.1103/PhysRevLett.110.268102} {\bibfield  {journal} {\bibinfo  {journal}
  {Phys. Rev. Lett.}\ }\textbf {\bibinfo {volume} {110}},\ \bibinfo {pages}
  {268102} (\bibinfo {year} {2013})}\BibitemShut {NoStop}%
\bibitem [{\citenamefont {Wioland}\ \emph {et~al.}(2016)\citenamefont
  {Wioland}, \citenamefont {Woodhouse}, \citenamefont {Dunkel},\ and\
  \citenamefont {Goldstein}}]{Wioland:2016kv}%
  \BibitemOpen
  \bibfield  {author} {\bibinfo {author} {\bibfnamefont {H.}~\bibnamefont
  {Wioland}}, \bibinfo {author} {\bibfnamefont {F.~G.}\ \bibnamefont
  {Woodhouse}}, \bibinfo {author} {\bibfnamefont {J.}~\bibnamefont {Dunkel}}, \
  and\ \bibinfo {author} {\bibfnamefont {R.~E.}\ \bibnamefont {Goldstein}},\
  }\bibfield  {title} {\enquote {\bibinfo {title} {Ferromagnetic and
  antiferromagnetic order in bacterial vortex lattices},}\ }\href {\doibase
  10.1038/nphys3607} {\bibfield  {journal} {\bibinfo  {journal} {Nat. Phys.}\
  }\textbf {\bibinfo {volume} {12}},\ \bibinfo {pages} {341} (\bibinfo {year}
  {2016})}\BibitemShut {NoStop}%
\bibitem [{\citenamefont {Guillamat}\ \emph {et~al.}()\citenamefont
  {Guillamat}, \citenamefont {Ign{\'e}s-Mullol},\ and\ \citenamefont
  {Sagu{\'e}s}}]{2015arXiv151103880G}%
  \BibitemOpen
  \bibfield  {author} {\bibinfo {author} {\bibfnamefont {P.}~\bibnamefont
  {Guillamat}}, \bibinfo {author} {\bibfnamefont {J.}~\bibnamefont
  {Ign{\'e}s-Mullol}}, \ and\ \bibinfo {author} {\bibfnamefont
  {F.}~\bibnamefont {Sagu{\'e}s}},\ }\bibfield  {title} {\enquote {\bibinfo
  {title} {Patterning active materials with addressable soft interfaces},}\
  }\href@noop {} {\ ,\ \bibinfo {pages} {3880}}\Eprint
  {http://arxiv.org/abs/arXiv:1511.03880v1} {arXiv:1511.03880v1} \BibitemShut
  {NoStop}%
\bibitem [{\citenamefont {Woodhouse}\ and\ \citenamefont
  {Goldstein}(2012)}]{Woodhouse:2012cl}%
  \BibitemOpen
  \bibfield  {author} {\bibinfo {author} {\bibfnamefont {F.~G.}\ \bibnamefont
  {Woodhouse}}\ and\ \bibinfo {author} {\bibfnamefont {R.~E.}\ \bibnamefont
  {Goldstein}},\ }\bibfield  {title} {\enquote {\bibinfo {title} {Spontaneous
  circulation of confined active suspensions},}\ }\href {\doibase
  10.1103/PhysRevLett.109.168105} {\bibfield  {journal} {\bibinfo  {journal}
  {Phys. Rev. Lett.}\ }\textbf {\bibinfo {volume} {109}},\ \bibinfo {pages}
  {168105} (\bibinfo {year} {2012})}\BibitemShut {NoStop}%
\bibitem [{\citenamefont {Segerer}\ \emph {et~al.}(2015)\citenamefont
  {Segerer}, \citenamefont {Th{\"u}roff}, \citenamefont {Piera~Alberola},
  \citenamefont {Frey},\ and\ \citenamefont {R{\"a}dler}}]{Segerer:2015js}%
  \BibitemOpen
  \bibfield  {author} {\bibinfo {author} {\bibfnamefont {F.~J.}\ \bibnamefont
  {Segerer}}, \bibinfo {author} {\bibfnamefont {F.}~\bibnamefont
  {Th{\"u}roff}}, \bibinfo {author} {\bibfnamefont {A.}~\bibnamefont
  {Piera~Alberola}}, \bibinfo {author} {\bibfnamefont {E.}~\bibnamefont
  {Frey}}, \ and\ \bibinfo {author} {\bibfnamefont {J.~O.}\ \bibnamefont
  {R{\"a}dler}},\ }\bibfield  {title} {\enquote {\bibinfo {title} {Emergence
  and persistence of collective cell migration on small circular
  micropatterns},}\ }\href {\doibase 10.1103/PhysRevLett.114.228102} {\bibfield
   {journal} {\bibinfo  {journal} {Phys. Rev. Lett.}\ }\textbf {\bibinfo
  {volume} {114}},\ \bibinfo {pages} {228102} (\bibinfo {year}
  {2015})}\BibitemShut {NoStop}%
\bibitem [{\citenamefont {Mohammad~Nejad}\ \emph {et~al.}(2014)\citenamefont
  {Mohammad~Nejad}, \citenamefont {Iannaccone}, \citenamefont {Rutherford},
  \citenamefont {Iannaccone},\ and\ \citenamefont
  {Foster}}]{MohammadNejad:2014he}%
  \BibitemOpen
  \bibfield  {author} {\bibinfo {author} {\bibfnamefont {T.}~\bibnamefont
  {Mohammad~Nejad}}, \bibinfo {author} {\bibfnamefont {S.}~\bibnamefont
  {Iannaccone}}, \bibinfo {author} {\bibfnamefont {W.}~\bibnamefont
  {Rutherford}}, \bibinfo {author} {\bibfnamefont {P.~M.}\ \bibnamefont
  {Iannaccone}}, \ and\ \bibinfo {author} {\bibfnamefont {C.~D.}\ \bibnamefont
  {Foster}},\ }\bibfield  {title} {\enquote {\bibinfo {title} {Mechanics and
  spiral formation in the rat cornea},}\ }\href {\doibase
  10.1007/s10237-014-0592-6} {\bibfield  {journal} {\bibinfo  {journal}
  {Biomech. Model. Mechanobiol.}\ }\textbf {\bibinfo {volume} {14}},\ \bibinfo
  {pages} {107} (\bibinfo {year} {2014})}\BibitemShut {NoStop}%
\bibitem [{\citenamefont {Whitfield}\ \emph {et~al.}(2014)\citenamefont
  {Whitfield}, \citenamefont {Marenduzzo}, \citenamefont {Voituriez},\ and\
  \citenamefont {Hawkins}}]{AWhitfield:2014in}%
  \BibitemOpen
  \bibfield  {author} {\bibinfo {author} {\bibfnamefont {C.~A.}\ \bibnamefont
  {Whitfield}}, \bibinfo {author} {\bibfnamefont {D.}~\bibnamefont
  {Marenduzzo}}, \bibinfo {author} {\bibfnamefont {R.}~\bibnamefont
  {Voituriez}}, \ and\ \bibinfo {author} {\bibfnamefont {R.~J.}\ \bibnamefont
  {Hawkins}},\ }\bibfield  {title} {\enquote {\bibinfo {title} {Active polar
  fluid flow in finite droplets},}\ }\href {\doibase
  10.1140/epje/i2014-14008-3} {\bibfield  {journal} {\bibinfo  {journal} {Eur.
  Phys. J. E}\ }\textbf {\bibinfo {volume} {37}},\ \bibinfo {pages} {8}
  (\bibinfo {year} {2014})}\BibitemShut {NoStop}%
\bibitem [{\citenamefont {Kumar}\ \emph {et~al.}(2014)\citenamefont {Kumar},
  \citenamefont {Maitra}, \citenamefont {Sumit}, \citenamefont {Ramaswamy},\
  and\ \citenamefont {Shivashankar}}]{Kumar:2014kv}%
  \BibitemOpen
  \bibfield  {author} {\bibinfo {author} {\bibfnamefont {A.}~\bibnamefont
  {Kumar}}, \bibinfo {author} {\bibfnamefont {A.}~\bibnamefont {Maitra}},
  \bibinfo {author} {\bibfnamefont {M.}~\bibnamefont {Sumit}}, \bibinfo
  {author} {\bibfnamefont {S.}~\bibnamefont {Ramaswamy}}, \ and\ \bibinfo
  {author} {\bibfnamefont {G.~V.}\ \bibnamefont {Shivashankar}},\ }\bibfield
  {title} {\enquote {\bibinfo {title} {Actomyosin contractility rotates the
  cell nucleus},}\ }\href {\doibase 10.1038/srep03781} {\bibfield  {journal}
  {\bibinfo  {journal} {Sci. Rep.}\ }\textbf {\bibinfo {volume} {4}},\ \bibinfo
  {pages} {3781} (\bibinfo {year} {2014})}\BibitemShut {NoStop}%
\bibitem [{\citenamefont {Simha}\ and\ \citenamefont
  {Ramaswamy}(2002)}]{AditiSimha:2002eg}%
  \BibitemOpen
  \bibfield  {author} {\bibinfo {author} {\bibfnamefont {A.~R.}\ \bibnamefont
  {Simha}}\ and\ \bibinfo {author} {\bibfnamefont {S.}~\bibnamefont
  {Ramaswamy}},\ }\bibfield  {title} {\enquote {\bibinfo {title} {Hydrodynamic
  fluctuations and instabilities in ordered suspensions of self-propelled
  particles},}\ }\href {\doibase 10.1103/PhysRevLett.89.058101} {\bibfield
  {journal} {\bibinfo  {journal} {Phys. Rev. Lett.}\ }\textbf {\bibinfo
  {volume} {89}},\ \bibinfo {pages} {058101} (\bibinfo {year}
  {2002})}\BibitemShut {NoStop}%
\bibitem [{\citenamefont {Sanchez}\ \emph {et~al.}(2013)\citenamefont
  {Sanchez}, \citenamefont {Chen}, \citenamefont {DeCamp}, \citenamefont
  {Heymann},\ and\ \citenamefont {Dogic}}]{Sanchez:2013gt}%
  \BibitemOpen
  \bibfield  {author} {\bibinfo {author} {\bibfnamefont {T.}~\bibnamefont
  {Sanchez}}, \bibinfo {author} {\bibfnamefont {D.~T.~N.}\ \bibnamefont
  {Chen}}, \bibinfo {author} {\bibfnamefont {S.~J.}\ \bibnamefont {DeCamp}},
  \bibinfo {author} {\bibfnamefont {M.}~\bibnamefont {Heymann}}, \ and\
  \bibinfo {author} {\bibfnamefont {Z.}~\bibnamefont {Dogic}},\ }\bibfield
  {title} {\enquote {\bibinfo {title} {Spontaneous motion in hierarchically
  assembled active matter},}\ }\href {\doibase 10.1038/nature11591} {\bibfield
  {journal} {\bibinfo  {journal} {Nature}\ }\textbf {\bibinfo {volume} {491}},\
  \bibinfo {pages} {431--434} (\bibinfo {year} {2013})}\BibitemShut {NoStop}%
\bibitem [{\citenamefont {Dunkel}\ \emph {et~al.}(2013)\citenamefont {Dunkel},
  \citenamefont {Heidenreich}, \citenamefont {Drescher}, \citenamefont
  {Wensink}, \citenamefont {B{\"a}r},\ and\ \citenamefont
  {Goldstein}}]{Dunkel:2013bm}%
  \BibitemOpen
  \bibfield  {author} {\bibinfo {author} {\bibfnamefont {J.}~\bibnamefont
  {Dunkel}}, \bibinfo {author} {\bibfnamefont {S.}~\bibnamefont {Heidenreich}},
  \bibinfo {author} {\bibfnamefont {K.}~\bibnamefont {Drescher}}, \bibinfo
  {author} {\bibfnamefont {H.~H.}\ \bibnamefont {Wensink}}, \bibinfo {author}
  {\bibfnamefont {M.}~\bibnamefont {B{\"a}r}}, \ and\ \bibinfo {author}
  {\bibfnamefont {R.~E.}\ \bibnamefont {Goldstein}},\ }\bibfield  {title}
  {\enquote {\bibinfo {title} {Fluid dynamics of bacterial turbulence},}\
  }\href {\doibase 10.1103/PhysRevLett.110.228102} {\bibfield  {journal}
  {\bibinfo  {journal} {Phys. Rev. Lett.}\ }\textbf {\bibinfo {volume} {110}},\
  \bibinfo {pages} {228102} (\bibinfo {year} {2013})}\BibitemShut {NoStop}%
\bibitem [{\citenamefont {Giomi}(2015{\natexlab{a}})}]{Giomi:2015fu}%
  \BibitemOpen
  \bibfield  {author} {\bibinfo {author} {\bibfnamefont {L.}~\bibnamefont
  {Giomi}},\ }\bibfield  {title} {\enquote {\bibinfo {title} {Geometry and
  topology of turbulence in active nematics},}\ }\href {\doibase
  10.1103/PhysRevX.5.031003} {\bibfield  {journal} {\bibinfo  {journal} {Phys.
  Rev. X}\ }\textbf {\bibinfo {volume} {5}},\ \bibinfo {pages} {031003}
  (\bibinfo {year} {2015}{\natexlab{a}})}\BibitemShut {NoStop}%
\bibitem [{\citenamefont {Adamer}\ \emph {et~al.}(2016)\citenamefont {Adamer},
  \citenamefont {Thampi}, \citenamefont {Yeomans},\ and\ \citenamefont
  {Doostmohammadi}}]{Adamer:2016bd}%
  \BibitemOpen
  \bibfield  {author} {\bibinfo {author} {\bibfnamefont {M.~F.}\ \bibnamefont
  {Adamer}}, \bibinfo {author} {\bibfnamefont {S.~P.}\ \bibnamefont {Thampi}},
  \bibinfo {author} {\bibfnamefont {J.~M.}\ \bibnamefont {Yeomans}}, \ and\
  \bibinfo {author} {\bibfnamefont {A.}~\bibnamefont {Doostmohammadi}},\
  }\bibfield  {title} {\enquote {\bibinfo {title} {Stabilization of active
  matter by flow-vortex lattices and defect ordering},}\ }\href {\doibase
  10.1038/ncomms10557} {\bibfield  {journal} {\bibinfo  {journal} {Nat.
  Commun.}\ }\textbf {\bibinfo {volume} {7}},\ \bibinfo {pages} {10557}
  (\bibinfo {year} {2016})}\BibitemShut {NoStop}%
\bibitem [{\citenamefont {Schaller}\ \emph {et~al.}(2010)\citenamefont
  {Schaller}, \citenamefont {Weber}, \citenamefont {Semmrich}, \citenamefont
  {Frey},\ and\ \citenamefont {Bausch}}]{Schaller:2010cq}%
  \BibitemOpen
  \bibfield  {author} {\bibinfo {author} {\bibfnamefont {V.}~\bibnamefont
  {Schaller}}, \bibinfo {author} {\bibfnamefont {C.}~\bibnamefont {Weber}},
  \bibinfo {author} {\bibfnamefont {C.}~\bibnamefont {Semmrich}}, \bibinfo
  {author} {\bibfnamefont {E.}~\bibnamefont {Frey}}, \ and\ \bibinfo {author}
  {\bibfnamefont {A.~R.}\ \bibnamefont {Bausch}},\ }\bibfield  {title}
  {\enquote {\bibinfo {title} {Polar patterns of driven filaments},}\ }\href
  {\doibase 10.1038/nature09312} {\bibfield  {journal} {\bibinfo  {journal}
  {Nature}\ }\textbf {\bibinfo {volume} {467}},\ \bibinfo {pages} {73}
  (\bibinfo {year} {2010})}\BibitemShut {NoStop}%
\bibitem [{\citenamefont {Sumino}\ \emph {et~al.}(2012)\citenamefont {Sumino},
  \citenamefont {Nagai}, \citenamefont {Shitaka}, \citenamefont {Tanaka},
  \citenamefont {Yoshikawa}, \citenamefont {Chat{\'e}},\ and\ \citenamefont
  {Oiwa}}]{Sumino:2012dw}%
  \BibitemOpen
  \bibfield  {author} {\bibinfo {author} {\bibfnamefont {Y.}~\bibnamefont
  {Sumino}}, \bibinfo {author} {\bibfnamefont {K.~H.}\ \bibnamefont {Nagai}},
  \bibinfo {author} {\bibfnamefont {Y.}~\bibnamefont {Shitaka}}, \bibinfo
  {author} {\bibfnamefont {D.}~\bibnamefont {Tanaka}}, \bibinfo {author}
  {\bibfnamefont {K.}~\bibnamefont {Yoshikawa}}, \bibinfo {author}
  {\bibfnamefont {H.}~\bibnamefont {Chat{\'e}}}, \ and\ \bibinfo {author}
  {\bibfnamefont {K.}~\bibnamefont {Oiwa}},\ }\bibfield  {title} {\enquote
  {\bibinfo {title} {Large-scale vortex lattice emerging from collectively
  moving microtubules},}\ }\href {\doibase 10.1038/nature10874} {\bibfield
  {journal} {\bibinfo  {journal} {Nature}\ }\textbf {\bibinfo {volume} {483}},\
  \bibinfo {pages} {448} (\bibinfo {year} {2012})}\BibitemShut {NoStop}%
\bibitem [{\citenamefont {Keber}\ \emph {et~al.}(2014)\citenamefont {Keber},
  \citenamefont {Loiseau}, \citenamefont {Sanchez}, \citenamefont {DeCamp},
  \citenamefont {Giomi}, \citenamefont {Bowick}, \citenamefont {Marchetti},
  \citenamefont {Dogic},\ and\ \citenamefont {Bausch}}]{Keber:2014fh}%
  \BibitemOpen
  \bibfield  {author} {\bibinfo {author} {\bibfnamefont {F.~C.}\ \bibnamefont
  {Keber}}, \bibinfo {author} {\bibfnamefont {E.}~\bibnamefont {Loiseau}},
  \bibinfo {author} {\bibfnamefont {T.}~\bibnamefont {Sanchez}}, \bibinfo
  {author} {\bibfnamefont {S.~J.}\ \bibnamefont {DeCamp}}, \bibinfo {author}
  {\bibfnamefont {L.}~\bibnamefont {Giomi}}, \bibinfo {author} {\bibfnamefont
  {M.~J.}\ \bibnamefont {Bowick}}, \bibinfo {author} {\bibfnamefont {M.~C.}\
  \bibnamefont {Marchetti}}, \bibinfo {author} {\bibfnamefont {Z.}~\bibnamefont
  {Dogic}}, \ and\ \bibinfo {author} {\bibfnamefont {A.~R.}\ \bibnamefont
  {Bausch}},\ }\bibfield  {title} {\enquote {\bibinfo {title} {Topology and
  dynamics of active nematic vesicles},}\ }\href {\doibase
  10.1126/science.1254784} {\bibfield  {journal} {\bibinfo  {journal}
  {Science}\ }\textbf {\bibinfo {volume} {345}},\ \bibinfo {pages} {1135}
  (\bibinfo {year} {2014})}\BibitemShut {NoStop}%
\bibitem [{\citenamefont {Hopf}(1989)}]{Hopf}%
  \BibitemOpen
  \bibfield  {author} {\bibinfo {author} {\bibfnamefont {H.}~\bibnamefont
  {Hopf}},\ }\href {https://books.google.co.uk/books?id=zNnw8NduHL0C} {\emph
  {\bibinfo {title} {Differential Geometry in the Large}}},\ Lecture Notes in
  Mathematics\ (\bibinfo  {publisher} {Springer Berlin Heidelberg},\ \bibinfo
  {year} {1989})\BibitemShut {NoStop}%
\bibitem [{\citenamefont {Giomi}\ \emph {et~al.}(2014)\citenamefont {Giomi},
  \citenamefont {Bowick}, \citenamefont {Mishra}, \citenamefont {Sknepnek},\
  and\ \citenamefont {Marchetti}}]{Giomi:2014ha}%
  \BibitemOpen
  \bibfield  {author} {\bibinfo {author} {\bibfnamefont {L.}~\bibnamefont
  {Giomi}}, \bibinfo {author} {\bibfnamefont {M.~J.}\ \bibnamefont {Bowick}},
  \bibinfo {author} {\bibfnamefont {P.}~\bibnamefont {Mishra}}, \bibinfo
  {author} {\bibfnamefont {R.}~\bibnamefont {Sknepnek}}, \ and\ \bibinfo
  {author} {\bibfnamefont {C.~M.}\ \bibnamefont {Marchetti}},\ }\bibfield
  {title} {\enquote {\bibinfo {title} {Defect dynamics in active nematics},}\
  }\href {\doibase 10.1098/rsta.2013.0365} {\bibfield  {journal} {\bibinfo
  {journal} {Phil. Trans. R. Soc. A}\ }\textbf {\bibinfo {volume} {372}},\
  \bibinfo {pages} {20130365} (\bibinfo {year} {2014})}\BibitemShut {NoStop}%
\bibitem [{\citenamefont {Sknepnek}\ and\ \citenamefont
  {Henkes}(2015)}]{Sknepnek:2015gm}%
  \BibitemOpen
  \bibfield  {author} {\bibinfo {author} {\bibfnamefont {R.}~\bibnamefont
  {Sknepnek}}\ and\ \bibinfo {author} {\bibfnamefont {S.}~\bibnamefont
  {Henkes}},\ }\bibfield  {title} {\enquote {\bibinfo {title} {Active swarms on
  a sphere},}\ }\href {\doibase 10.1103/PhysRevE.91.022306} {\bibfield
  {journal} {\bibinfo  {journal} {Phys. Rev. E}\ }\textbf {\bibinfo {volume}
  {91}},\ \bibinfo {pages} {022306} (\bibinfo {year} {2015})}\BibitemShut
  {NoStop}%
\bibitem [{\citenamefont {Oron}\ \emph {et~al.}(1997)\citenamefont {Oron},
  \citenamefont {Davis},\ and\ \citenamefont {Bankoff}}]{1997RvMP...69..931O}%
  \BibitemOpen
  \bibfield  {author} {\bibinfo {author} {\bibfnamefont {A.}~\bibnamefont
  {Oron}}, \bibinfo {author} {\bibfnamefont {S.~H.}\ \bibnamefont {Davis}}, \
  and\ \bibinfo {author} {\bibfnamefont {S.~G.}\ \bibnamefont {Bankoff}},\
  }\bibfield  {title} {\enquote {\bibinfo {title} {Long-scale evolution of thin
  liquid films},}\ }\href {\doibase 10.1103/RevModPhys.69.931} {\bibfield
  {journal} {\bibinfo  {journal} {Rev. Mod. Phys.}\ }\textbf {\bibinfo {volume}
  {69}},\ \bibinfo {pages} {931} (\bibinfo {year} {1997})}\BibitemShut
  {NoStop}%
\bibitem [{\citenamefont {Takagi}\ and\ \citenamefont
  {Huppert}(2010)}]{TAKAGI:2010fi}%
  \BibitemOpen
  \bibfield  {author} {\bibinfo {author} {\bibfnamefont {D.}~\bibnamefont
  {Takagi}}\ and\ \bibinfo {author} {\bibfnamefont {H.~E.}\ \bibnamefont
  {Huppert}},\ }\bibfield  {title} {\enquote {\bibinfo {title} {Flow and
  instability of thin films on a cylinder and sphere},}\ }\href {\doibase
  10.1017/S0022112009993818} {\bibfield  {journal} {\bibinfo  {journal} {J.
  Fluid Mech.}\ }\textbf {\bibinfo {volume} {647}},\ \bibinfo {pages} {221}
  (\bibinfo {year} {2010})}\BibitemShut {NoStop}%
\bibitem [{\citenamefont {Sankararaman}\ and\ \citenamefont
  {Ramaswamy}(2009)}]{Sankararaman:2009bx}%
  \BibitemOpen
  \bibfield  {author} {\bibinfo {author} {\bibfnamefont {S.}~\bibnamefont
  {Sankararaman}}\ and\ \bibinfo {author} {\bibfnamefont {S.}~\bibnamefont
  {Ramaswamy}},\ }\bibfield  {title} {\enquote {\bibinfo {title} {Instabilities
  and waves in thin films of living fluids},}\ }\href {\doibase
  10.1103/PhysRevLett.102.118107} {\bibfield  {journal} {\bibinfo  {journal}
  {Phys. Rev. Lett.}\ }\textbf {\bibinfo {volume} {102}},\ \bibinfo {pages}
  {118107} (\bibinfo {year} {2009})}\BibitemShut {NoStop}%
\bibitem [{\citenamefont {Joanny}\ and\ \citenamefont
  {Ramaswamy}(2012)}]{Joanny:2012dx}%
  \BibitemOpen
  \bibfield  {author} {\bibinfo {author} {\bibfnamefont {J.~F.}\ \bibnamefont
  {Joanny}}\ and\ \bibinfo {author} {\bibfnamefont {S.}~\bibnamefont
  {Ramaswamy}},\ }\bibfield  {title} {\enquote {\bibinfo {title} {A drop of
  active matter},}\ }\href {\doibase 10.1017/jfm.2012.131} {\bibfield
  {journal} {\bibinfo  {journal} {J. Fluid Mech.}\ }\textbf {\bibinfo {volume}
  {705}},\ \bibinfo {pages} {46} (\bibinfo {year} {2012})}\BibitemShut
  {NoStop}%
\bibitem [{\citenamefont {Khoromskaia}\ and\ \citenamefont
  {Alexander}(2015)}]{Khoromskaia:2015ec}%
  \BibitemOpen
  \bibfield  {author} {\bibinfo {author} {\bibfnamefont {D.}~\bibnamefont
  {Khoromskaia}}\ and\ \bibinfo {author} {\bibfnamefont {G.~P.}\ \bibnamefont
  {Alexander}},\ }\bibfield  {title} {\enquote {\bibinfo {title} {Motility of
  active fluid drops on surfaces},}\ }\href {\doibase
  10.1103/PhysRevE.92.062311} {\bibfield  {journal} {\bibinfo  {journal} {Phys.
  Rev. E}\ }\textbf {\bibinfo {volume} {92}} (\bibinfo {year} {2015}),\
  10.1103/PhysRevE.92.062311}\BibitemShut {NoStop}%
\bibitem [{\citenamefont {Chaikin}\ and\ \citenamefont
  {Lubensky}(1995)}]{Chaikin}%
  \BibitemOpen
  \bibfield  {author} {\bibinfo {author} {\bibfnamefont {P.~M.}\ \bibnamefont
  {Chaikin}}\ and\ \bibinfo {author} {\bibfnamefont {T.~C.}\ \bibnamefont
  {Lubensky}},\ }\href {\doibase 10.1007/BF02179565} {\emph {\bibinfo {title}
  {Principles of Condensed Matter Physics}}}\ (\bibinfo  {publisher} {Cambridge
  Univ Press},\ \bibinfo {year} {1995})\BibitemShut {NoStop}%
\bibitem [{\citenamefont {Lubensky}\ and\ \citenamefont
  {Prost}(1992)}]{Lubensky:1992bn}%
  \BibitemOpen
  \bibfield  {author} {\bibinfo {author} {\bibfnamefont {T~C}\ \bibnamefont
  {Lubensky}}\ and\ \bibinfo {author} {\bibfnamefont {J}~\bibnamefont
  {Prost}},\ }\bibfield  {title} {\enquote {\bibinfo {title} {{Orientational
  order and vesicle shape}},}\ }\href {\doibase 10.1051/jp2:1992133>}
  {\bibfield  {journal} {\bibinfo  {journal} {Journal de Physique II}\ }
  (\bibinfo {year} {1992}),\ 10.1051/jp2:1992133>}\BibitemShut {NoStop}%
\bibitem [{\citenamefont {Pismen}(2013)}]{Pismen:2013ie}%
  \BibitemOpen
  \bibfield  {author} {\bibinfo {author} {\bibfnamefont {L.~M.}\ \bibnamefont
  {Pismen}},\ }\bibfield  {title} {\enquote {\bibinfo {title} {Dynamics of
  defects in an active nematic layer},}\ }\href {\doibase
  10.1103/PhysRevE.88.050502} {\bibfield  {journal} {\bibinfo  {journal} {Phys.
  Rev. E}\ }\textbf {\bibinfo {volume} {88}},\ \bibinfo {pages} {050502}
  (\bibinfo {year} {2013})}\BibitemShut {NoStop}%
\bibitem [{\citenamefont {Vitelli}\ and\ \citenamefont
  {Nelson}(2006)}]{Vitelli:2006ba}%
  \BibitemOpen
  \bibfield  {author} {\bibinfo {author} {\bibfnamefont {V.}~\bibnamefont
  {Vitelli}}\ and\ \bibinfo {author} {\bibfnamefont {D.~R.}\ \bibnamefont
  {Nelson}},\ }\bibfield  {title} {\enquote {\bibinfo {title} {Nematic textures
  in spherical shells},}\ }\href {\doibase 10.1103/PhysRevE.74.021711}
  {\bibfield  {journal} {\bibinfo  {journal} {Phys. Rev. E}\ }\textbf {\bibinfo
  {volume} {74}},\ \bibinfo {pages} {021711} (\bibinfo {year}
  {2006})}\BibitemShut {NoStop}%
\bibitem [{\citenamefont {Kruse}\ \emph {et~al.}(2004)\citenamefont {Kruse},
  \citenamefont {Joanny}, \citenamefont {J{\"u}licher}, \citenamefont {Prost},\
  and\ \citenamefont {Sekimoto}}]{Kruse:2004il}%
  \BibitemOpen
  \bibfield  {author} {\bibinfo {author} {\bibfnamefont {K.}~\bibnamefont
  {Kruse}}, \bibinfo {author} {\bibfnamefont {J.~F.}\ \bibnamefont {Joanny}},
  \bibinfo {author} {\bibfnamefont {F.}~\bibnamefont {J{\"u}licher}}, \bibinfo
  {author} {\bibfnamefont {J.}~\bibnamefont {Prost}}, \ and\ \bibinfo {author}
  {\bibfnamefont {K.}~\bibnamefont {Sekimoto}},\ }\bibfield  {title} {\enquote
  {\bibinfo {title} {Asters, vortices, and rotating spirals in active gels of
  polar filaments},}\ }\href {\doibase 10.1103/PhysRevLett.92.078101}
  {\bibfield  {journal} {\bibinfo  {journal} {Phys. Rev. Lett.}\ }\textbf
  {\bibinfo {volume} {92}},\ \bibinfo {pages} {078101} (\bibinfo {year}
  {2004})}\BibitemShut {NoStop}%
\bibitem [{\citenamefont {Nelson}(2002)}]{Nelson:2002dd}%
  \BibitemOpen
  \bibfield  {author} {\bibinfo {author} {\bibfnamefont {David~R}\ \bibnamefont
  {Nelson}},\ }\bibfield  {title} {\enquote {\bibinfo {title} {{Toward a
  Tetravalent Chemistry of Colloids}},}\ }\href {\doibase 10.1021/nl0202096}
  {\bibfield  {journal} {\bibinfo  {journal} {Nano Letters}\ }\textbf {\bibinfo
  {volume} {2}},\ \bibinfo {pages} {1125--1129} (\bibinfo {year}
  {2002})}\BibitemShut {NoStop}%
\bibitem [{\citenamefont {Nedelec}\ \emph {et~al.}(1997)\citenamefont
  {Nedelec}, \citenamefont {Surrey}, \citenamefont {Maggs},\ and\ \citenamefont
  {Leibler}}]{1997Natur.389..305N}%
  \BibitemOpen
  \bibfield  {author} {\bibinfo {author} {\bibfnamefont {F.~J.}\ \bibnamefont
  {Nedelec}}, \bibinfo {author} {\bibfnamefont {T.}~\bibnamefont {Surrey}},
  \bibinfo {author} {\bibfnamefont {A.~C.}\ \bibnamefont {Maggs}}, \ and\
  \bibinfo {author} {\bibfnamefont {S.}~\bibnamefont {Leibler}},\ }\bibfield
  {title} {\enquote {\bibinfo {title} {Self-organization of microtubules and
  motors},}\ }\href {\doibase 10.1038/38532} {\bibfield  {journal} {\bibinfo
  {journal} {Nature}\ }\textbf {\bibinfo {volume} {389}},\ \bibinfo {pages}
  {305} (\bibinfo {year} {1997})}\BibitemShut {NoStop}%
\bibitem [{\citenamefont {Giomi}(2015{\natexlab{b}})}]{2015PhRvX...5c1003G}%
  \BibitemOpen
  \bibfield  {author} {\bibinfo {author} {\bibfnamefont {Luca}\ \bibnamefont
  {Giomi}},\ }\bibfield  {title} {\enquote {\bibinfo {title} {{Geometry and
  Topology of Turbulence in Active Nematics}},}\ }\href {\doibase
  10.1103/PhysRevX.5.031003} {\bibfield  {journal} {\bibinfo  {journal}
  {Physical Review X}\ }\textbf {\bibinfo {volume} {5}},\ \bibinfo {pages}
  {031003} (\bibinfo {year} {2015}{\natexlab{b}})}\BibitemShut {NoStop}%
\bibitem [{\citenamefont {Selinger}\ \emph {et~al.}(2011)\citenamefont
  {Selinger}, \citenamefont {Konya}, \citenamefont {Travesset},\ and\
  \citenamefont {Selinger}}]{Selinger:2011ky}%
  \BibitemOpen
  \bibfield  {author} {\bibinfo {author} {\bibfnamefont {Robin L~Blumberg}\
  \bibnamefont {Selinger}}, \bibinfo {author} {\bibfnamefont {Andrew}\
  \bibnamefont {Konya}}, \bibinfo {author} {\bibfnamefont {Alex}\ \bibnamefont
  {Travesset}}, \ and\ \bibinfo {author} {\bibfnamefont {Jonathan~V}\
  \bibnamefont {Selinger}},\ }\bibfield  {title} {\enquote {\bibinfo {title}
  {{Monte Carlo Studies of the XY Model on Two-Dimensional Curved Surfaces}},}\
  }\href {\doibase 10.1021/jp205128g} {\bibfield  {journal} {\bibinfo
  {journal} {The Journal of Physical Chemistry B}\ }\textbf {\bibinfo {volume}
  {115}},\ \bibinfo {pages} {13989--13993} (\bibinfo {year}
  {2011})}\BibitemShut {NoStop}%
\bibitem [{\citenamefont {Stone}\ and\ \citenamefont
  {Samuel}(1996)}]{Stone:1996ir}%
  \BibitemOpen
  \bibfield  {author} {\bibinfo {author} {\bibfnamefont {H.~A.}\ \bibnamefont
  {Stone}}\ and\ \bibinfo {author} {\bibfnamefont {A.D.T.}\ \bibnamefont
  {Samuel}},\ }\bibfield  {title} {\enquote {\bibinfo {title} {Propulsion of
  microorganisms by surface distortions},}\ }\href {\doibase
  10.1103/PhysRevLett.77.4102} {\bibfield  {journal} {\bibinfo  {journal}
  {Phys. Rev. Lett.}\ }\textbf {\bibinfo {volume} {77}},\ \bibinfo {pages}
  {4102} (\bibinfo {year} {1996})}\BibitemShut {NoStop}%
\bibitem [{\citenamefont {Vlahovska}(2016)}]{10.1039/9781782628491-00313}%
  \BibitemOpen
  \bibfield  {author} {\bibinfo {author} {\bibfnamefont {P.~M.}\ \bibnamefont
  {Vlahovska}},\ }\bibfield  {title} {\enquote {\bibinfo {title} {Dynamics of
  membrane-bound particles: Capsules and vesicles},}\ }in\ \href {\doibase
  10.1039/9781782628491-00313} {\emph {\bibinfo {booktitle} {Fluid-Structure
  Interactions in Low-Reynolds-Number Flows}}}\ (\bibinfo  {publisher} {The
  Royal Society of Chemistry},\ \bibinfo {year} {2016})\ pp.\ \bibinfo {pages}
  {313--346}\BibitemShut {NoStop}%
\bibitem [{\citenamefont {Khoromskaia}\ \emph {et~al.}()\citenamefont
  {Khoromskaia}, \citenamefont {Whitfield},\ and\ \citenamefont
  {Alexander}}]{futurepaper}%
  \BibitemOpen
  \bibfield  {author} {\bibinfo {author} {\bibfnamefont {D.}~\bibnamefont
  {Khoromskaia}}, \bibinfo {author} {\bibfnamefont {C.~A.}\ \bibnamefont
  {Whitfield}}, \ and\ \bibinfo {author} {\bibfnamefont {G.~P.}\ \bibnamefont
  {Alexander}},\ }\href@noop {} {\bibinfo  {journal} {in preparation}\
  }\BibitemShut {NoStop}%
\bibitem [{\citenamefont {Pairam}\ \emph {et~al.}(2013)\citenamefont {Pairam},
  \citenamefont {Vallamkondu}, \citenamefont {Koning}, \citenamefont {van
  Zuiden}, \citenamefont {Ellis}, \citenamefont {Bates}, \citenamefont
  {Vitelli},\ and\ \citenamefont {Fernandez-Nieves}}]{Pairam04062013}%
  \BibitemOpen
\bibfield  {journal} {  }\bibfield  {author} {\bibinfo {author} {\bibfnamefont
  {E.}~\bibnamefont {Pairam}}, \bibinfo {author} {\bibfnamefont
  {J.}~\bibnamefont {Vallamkondu}}, \bibinfo {author} {\bibfnamefont
  {V.}~\bibnamefont {Koning}}, \bibinfo {author} {\bibfnamefont {B.~C.}\
  \bibnamefont {van Zuiden}}, \bibinfo {author} {\bibfnamefont {P.~W.}\
  \bibnamefont {Ellis}}, \bibinfo {author} {\bibfnamefont {M.~A.}\ \bibnamefont
  {Bates}}, \bibinfo {author} {\bibfnamefont {V.}~\bibnamefont {Vitelli}}, \
  and\ \bibinfo {author} {\bibfnamefont {A.}~\bibnamefont {Fernandez-Nieves}},\
  }\bibfield  {title} {\enquote {\bibinfo {title} {Stable nematic droplets with
  handles},}\ }\href {\doibase 10.1073/pnas.1221380110} {\bibfield  {journal}
  {\bibinfo  {journal} {Proc. Natl. Acad. Sci. U.S.A.}\ }\textbf {\bibinfo
  {volume} {110}},\ \bibinfo {pages} {9295} (\bibinfo {year}
  {2013})}\BibitemShut {NoStop}%
\end{thebibliography}%

\end{document}